\documentclass{emulateapj}
\usepackage[optb]{optional}

\usepackage{ifthen}

\newboolean{AAorApJ}

\setboolean{AAorApJ}{false}

\ifthenelse{\boolean{AAorApJ}}
{\usepackage{natbib}
  \bibpunct{(}{)}{;}{a}{}{,} % to follow the A&A style
}
{\usepackage{natbib}}

\usepackage{graphicx}

\ifthenelse{\boolean{AAorApJ}}
{}
{\usepackage{amsmath,amssymb,amsthm,amsfonts,amscd}}
\ifthenelse{\boolean{AAorApJ}}
{\usepackage{txfonts}  }
{\usepackage{apjfonts} }

\begin{document}

%\journalinfo{The Astrophysical Journal}
\slugcomment{Received 2009 Month XX; accepted 2009 Month XX}

\shortauthors{Hammer et al.}
\shorttitle{Three-dimensional supernova explosion models}

%  Definitions
\def\ergs{\,erg\,s$^{-1}$}
\def\pccc{\,pc\,cm$^{-3}$}
\def\rlum{\,mJy\,kpc$^2$}

% Objects
\def\psr{PSR~J1747--2809}
\def\snr{G0.9+0.1}
\def\cxou{CXOU~J174722.8--280915}
\def\psrb{PSR~J1833--1034}
\def\snrb{G21.5--0.9}

\title{Three-dimensional simulations of mixing instabilities in
  supernova explosions}

\author{N.J.~Hammer,\altaffilmark{1}
  H.-Th.~Janka,\altaffilmark{1}
  and E.~M\"uller\altaffilmark{1}
}

\altaffiltext{1}{Max-Planck-Institut f\"ur Astrophysik, 
  Karl-Schwarzschild-Str.~1, D-85748 Garching, Germany;
  corresponding author: thj@mpa-garching.mpg.de}

\begin{abstract}
We present the first three-dimensional (3D) simulations of the
large-scale mixing that takes place in the shock-heated stellar layers
ejected in the explosion of a 15.5$\,M_\odot$ blue supergiant
star. The blast is initiated and powered by neutrino-energy deposition
behind the stalled shock by means of choosing sufficiently high
neutrino luminosities from the contracting, nascent neutron star, whose
high-density core is excised and replaced by a retreating inner grid
boundary. The outgoing supernova shock is followed beyond its breakout
from the stellar surface more than two hours after the core
collapse. Violent convective overturn in the post-shock layer causes
the explosion to start with significant large-scale asphericity, which
acts as a trigger of the growth of Rayleigh-Taylor instabilities at
the composition interfaces of the exploding star. Despite the absence
of a strong Richtmyer-Meshkov instability at the H/He interface, which
only a largely deformed shock could instigate, deep inward mixing of
hydrogen is found as well as fast-moving, metal-rich clumps
penetrating with high velocities far into the hydrogen envelope of the
star as observed, for example, in the case of Supernova~1987A.  Also
individual clumps containing a sizeable fraction of the ejected
iron-group elements (up to several $10^{-3}\,M_\odot$) are obtained in 
some models. The metal core of the progenitor is partially turned over
with nickel-dominated fingers overtaking oxygen-rich bullets and both
nickel and oxygen moving well ahead of the material from the carbon 
layer. Comparing with corresponding two-dimensional (axially
symmetric; 2D) calculations, we determine the growth of the 
Rayleigh-Taylor fingers to be faster, the deceleration of the
dense metal-carrying clumps in the helium and hydrogen layers to be
reduced, the asymptotic clump velocities in the hydrogen shell to
be higher (up to $\sim$4500$\,$km$\,$s$^{-1}$ for the considered
progenitor and an explosion energy of $10^{51}\,$ergs, instead of 
$\la$2000$\,$km$\,$s$^{-1}$ in 2D), and the outward radial mixing of
heavy elements and inward mixing of hydrogen to be more efficient in 3D
than in 2D. We present a simple argument that explains these results as
a consequence of the different action of drag forces on moving objects
in the two geometries.
\end{abstract}


\section{Introduction} \label{sec:intro}
Besides an experimental confirmation of a core collapse event through
the detection of supernova neutrinos \citep{hirata:1987a}, the second
most important insight provided by observations of Supernova 1987A was
the occurrence of large-scale non-radial flow and extensive mixing of
chemical species in the envelope of the progenitor star during the
explosion \citep[see, e.g.,][]{arnett:1989a}. While SN\,1987A still
remains the most prominent and thoroughly-observed example,
observations of many other core-collapse supernovae (SNe) have
meanwhile provided ample evidence that large-scale extensive mixing
seems to occur generically in such events \citep[see,
  e.g.,][]{wang:2008a}. In particular, the modeling of the light curve
of SN\,1987A using 1D radiation hydrodynamic calculations requires a
large amount of mixing of Ni outward to the H-He interface and of H
inward into the He-core \citep{woosley:1988d, shigeyama:1990b,
  blinnikov:2000a, utrobin:2004a}.  Moreover, the asymmetry of iron
and nickel lines in SN\,1987 can be explained, if Ni is concentrated
into many high-velocity bullets \citep{spyromilio:1990a, li:1993a}.
On the other hand, the Bochum event, i.e., the sudden development of
fine-structure in the $H_\alpha$ line about two weeks after the
explosion \citep{hanuschik:1988a}, implies that a high velocity ($\sim
4700\,$km/s) clump of $^{56}$Ni with a mass of $~ 10^{-3} M_\odot$ was
ejected into the far hemisphere of SN\,1987 \citep{utrobin:1995a}.

Observations of near-infrared He\,I lines from Type\,II SNe between 50
and 100 days after core collapse imply mixing of $^{56}$Ni into the
hydrogen envelope \citep{fassia:1998a, fassia:1999a}. These authors
further conclude that those lines are formed in a clumpy
environment. Dense knots, indications of ejecta shrapnels, and
filaments seen in supernova remnants by HST in the visible
\citep{blair:2000a, fesen:2006a} and by ROSAT, Chandra, and XMM in
X-rays \citep{aschenbach:1995a, hughes:2000a, miceli:2008a} also
provide strong evidence that mixing and perhaps even fragmentation is
a common process in supernova explosions.  Spectropolarimetric
observations of Type\,II-P and Type\,IIn SNe at late epochs, when the
hydrogen envelope starts thinning, reveal strong evidence for a
globally highly aspherical distribution of the inner ejecta
\citep{fassia:1998a, leonard:2001a, leonard:2006a}.  Similar results
are obtained from spectropolarimetric observations of Type\,Ib/c SNe
\citep{kawabata:2002a, wang:2003a, maund:2007a, modjaz:2008a,
  wang:2008a}.  Finally, studying the nebula spectra of SN\,Ic 2002ap
by the means of synthetic spectra, \citet{mazzali:2007a} found
evidence for an oxygen-rich inner core and $^{56}$Ni at high
velocities, suggesting a highly aspherical explosion especially in the
inner parts.

Two-dimensional hydrodynamic simulations of non-radial flow and mixing
in the stellar envelopes of core collapse supernova progenitors have
been performed by several groups. The first simulations were started
by artificially seeding Rayleigh-Taylor instabilities (RTIs) in the
mantle and envelope of the progenitor and following their evolution
until shock breakout from the stellar surface \citep{arnett:1989a,
  fryxell:1991a, mueller:1991a, mueller:1991b, hachisu:1990a,
  herant:1991a, hachisu:1992a, herant:1992a}. More recent simulations
consistently connect the seed asymmetries arising from convective flow
in the neutrino-heated bubble, and by the standing accretion shock
instability (SASI) in the ``supernova engine'' during the first second
of the explosion \citep[see, e.g.,][]{janka:2007a} to the later
Rayleigh-Taylor and Richtmyer-Meshkov instabilities after shock
passage through the outer stellar layers with application to SN\,1987A
\citep{kifonidis:2003a, kifonidis:2006a}. Rayleigh-Taylor induced
mixing and the amount of fallback occurring during artificially
triggered (piston model), and initially spherically symmetric
supernova explosions of zero- and solar-metallicity 15 and
$25\,M_\odot$ stars were studied by \citet{joggerst:2009a}. In all
models considered by them the velocities of the heavy elements were
less than those observed in SN\,1987A. \citet{couch:2009a} explored
the observational characteristics of jet-driven supernovae in a red
supergiant progenitor injecting $10^{51}\,$ergs of kinetic energy and
varying the characteristics of the assumed bipolar outflows (jets).
Their simulations show the development of fluid instabilities that
produce helium clumps in the hydrogen envelope, and nickel-rich jets
that may account for the non-spherical excitation and substructure of
spectral lines.

Only a few attempts to 3D simulations have been made up to now.
\citet{nagasawa:1988a} simulated the adiabatic point explosion of an
$n = 3$ polytrope using a 3D smoothed particle hydrodynamics (SPH)
code, and claimed to have found pronounced Rayleigh-Taylor
instabilities in that event.  However, this claim was neither
confirmed by the 3D simulations of \citet{mueller:1989a} using a 3D
hydrodynamics code based on a high-resolution shock-capturing
finite-volume method, nor by \citet{benz:1990a} who also used a 3D SPH
code.  Both studies found only a weak instability, i.e., no extensive
clumping and mixing, consistent with linear stability
analysis. Subsequently, 3D simulations by \cite{mueller:1991b} and
\citet{yamada:1990a} using relatively coarse resolution and/or
considering only a wedge-shaped fraction of the star showed that seed
perturbations grow strongly at the He-H interface when a realistic
presupernova star model is used instead of an $n=3$ polytrope. Later
\citet{kane:2000a} studied the difference in the growth of 2D versus
3D single-mode perturbations at the He-H and O-He interfaces of
SN\,1987A. They found that the 3D single-mode perturbation grows about
30\% faster than a corresponding 2D one. Although the difference
between 2D and 3D predicted by their simulations is not enough to
account for the difference between observed $^{56}$Ni velocities in SN
1987A and the results of previous 2D simulations of SN 1987A, their
results suggest that non-radial flow and mixing in supernova envelopes
in full 3D is noticeably different from that predicted by 2D
simulations. This finding is also supported by laser experiments and
comprehensive Rayleigh-Taylor growth simulations
\citep{anuchina:2004a, cabot:2006a, remington:2006a}. The only other
3D simulations we are aware of are those published by
\citet{noro:2002a, noro:2004a} in two short conference
proceedings. These authors simulated the propagation of a 1D shock
wave resulting from an artificially triggered spherical explosion of
$10^{51}\,$ergs until it reaches the He-H interface of a SN\,1987A
progenitor model computed by \citet{shigeyama:1990b}. Then they
perturbed the flow inside the shock radius and followed the growth of
sinusoidal velocity perturbations with a 3D adaptive mesh hydro-code
on a grid with an effective resolution of up to $4096^3$ zones.  For
an initial perturbation amplitude of 10\%, fingers of heavy elements
penetrate into the hydrogen envelope with large velocities.

Here we simulate for the first time, starting with 3D models for the
beginning of neutrino-powered explosions from \citet{scheck:2007a},
the evolution of a supernova blast in 3D from the first hundreds of
milliseconds to a time three hours later when the shock has broken out
of a blue progenitor star. We find that not only the growth of
Rayleigh-Taylor instabilities is different in 3D as concluded from
other works (see above), but also that the propagation of ``clumps''
in 3D is different from the 2D case. In particular the latter effect
is of large quantitative importance, because we demonstrate here that
it determines the long-time evolution and in the end the extent of
large-scale radial mixing in core-collapse supernovae.  Quantitatively
meaningful simulations of observable explosion asymmetries therefore
require modeling in three spatial dimensions.

The paper is organized as follows. In Sect.~2 we describe the
computational setup and the initial model used in our simulations.
Our results are discussed in Sect.~3 where we particularly address the
dynamical evolution and the radial mixing of chemical elements. In
Sect.~4 we present a simple analytical model that explains the
differences found for the clump propagation between 2D and 3D quite
well. Finally, we summarize our findings in Sect.~5, and draw some
conclusions.

%%%%%%%%%%%%%%%%%%%%%%%%%%%%%%%%%%%%%%%%%%%%%%%%%%%%%%%%
%% FIGURE 1
%%%%%%%%%%%%%%%%%%%%%%%%%%%%%%%%%%%%%%%%%%%%%%%%%%%%%%%%
\begin{figure}[tpb!]
\begin{center}
\plotone{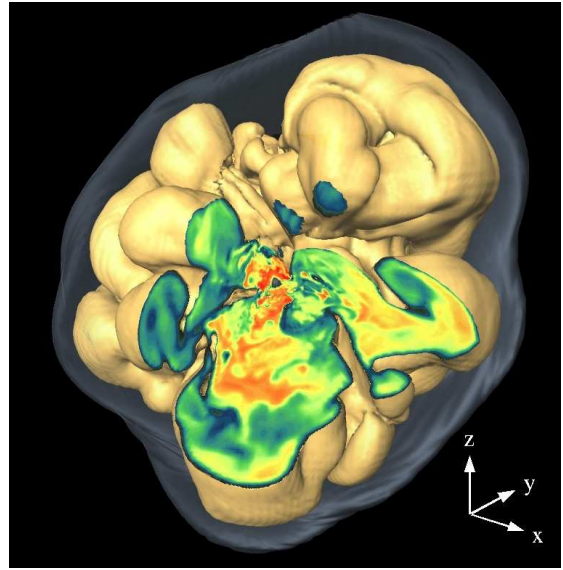}
\caption{\label{fig:inimod} 
  Three-dimensional, neutrino-driven explosion simulation at about
  0.5$\,$s after core bounce \citep{scheck:2007a}. The model is used
  as initial data for our studies. The outermost, bluish, nearly
  transparent surface is the supernova shock with an average radius of
  1900$\,$km. The bright bumpy surface is the interface between
  rising, high-entropy neutrino-heated matter and lower-entropy,
  accreted post-shock gas. An octant is cut out to show the entropy
  distribution in the rising Rayleigh-Taylor bubbles and cooler
  downflows (the entropies per nucleon vary between about 10 and
  21$\,k_{\mathrm{B}}$ from blue to yellow and red).  The neutron star
  is visible as a dark grey surface near the center.}
\end{center}
\end{figure}
%%%%%%%%%%%%%%%%%%%%%%%%%%%%%%%%%%%%%%%%%%%%%%%%%%%%%%%%

\section{Simulation setup}

For our simulations we used the explicit finite-volume Eulerian
multi-fluid hydrodynamics code, PROMETHEUS, developed by Bruce Fryxell
and Ewald M\"uller \citep{fryxell:1991a, mueller:1991a,
mueller:1991b}.  The code integrates the equations of multidimensional
hydrodynamics to second order accuracy in space and time using the
dimensional splitting approach of \citet{Strang:1968}, the PPM
reconstruction scheme \citep{ColellaWoodward:1984}, and a Riemann
solver for real gases according to \citet{ColellaGlaz:1984}.
Multi-fluid flows (involving the advection of, e.g., several nuclear
species; see below) are treated by means of a set of additional
continuity equations using the Consistent Multi-fluid Advection (CMA)
method of \citet{PlewaMueller:1999}.

\subsection{Computational grid}
We use a computational grid in spherical polar coordinates consisting
of $1200 \times 180 \times 360$ zones in $r$, $\theta$, and
$\phi$-direction. The equidistant angular grid has a resolution of
$0.935^{\circ}$ in $\theta$-direction and $1^{\circ}$ in
$\phi$-direction, respectively, covering the whole sphere except for a
(double) cone with a half opening angle of $5.8^{\circ}$ around the
symmetry axis of the coordinate system (i.e., $0.0325\,\pi \leq \theta
\leq 0.9675\,\pi$, and $0 \leq \phi \leq 2\pi$).  Reflective boundary
conditions are imposed in $\theta$-direction, and periodic ones in
$\phi$-direction. The clipping of the computational grid around the
symmetry axis avoids a too restrictive CFL time step condition. We
have not observed any numerical artifacts as consequences of this
grid constraint. 

The radial mesh is logarithmically spaced between a time-dependent
inner boundary, initially located at a radius of 200\,km, and a fixed
outer boundary at $3.9 \times 10^{12}\,$cm.  It has a maximum
resolution of 2\,km at the inner boundary, and a resolution of $4
\times 10^{10}\,$cm at the outer one (corresponding to a roughly
constant relative radial resolution $\Delta r/ r \approx 1\%$). We
allow for free outflow at the outer boundary, and impose a reflective
boundary condition at the inner edge of the radial grid.

During the simulation we move (approximately every 100th timestep) the
inner radial boundary to larger and larger radii to relax the CFL
condition.  This cutting of the computational grid reduces the number
of (the logarithmically spaced) radial zones from 1200 at the
beginning to about 400 towards the end of the simulation, but involves
only the innermost few percent of the initial envelope mass. We
convinced ourselves by means of 2D test calculations that this removal
of mass has no effect on the dynamics and mixing occurring at larger
radii (the cut radius exceeds at no time a few percent of the radius
of the Rayleigh-Taylor finger tips).

\subsection{Initial model}
The initial model used for our 3D simulations is based on 3D
hydrodynamic explosion models of \citet{scheck:2007a}, who followed
the onset and early development of neutrino-driven explosions in 3D
with the same code, numerical setup, and input physics as described in
all details for 1D and 2D simulations in \citet{scheck:2006a}.  These
runs were started from 1D stellar collapse models a few milliseconds
after core bounce. The explosion was initiated and powered by the
energy input of neutrinos in the postshock layer, where vigorous
convective activity was present and supported the revival of the
stalled prompt shock. A defined value of the explosion energy was
obtained by suitably choosing the magnitude of the neutrino
luminosities that were imposed at the retreating inner grid boundary,
which replaced the excised, contracting dense core of the forming
neutron star.

The stellar progenitor considered here is a spherically symmetric
model of a $15.5\,M_\odot$ blue supergiant with a radius of $4\times
10^{12}\,$cm \citep[Woosley, private communication; see also Kifonidis
  et al.\ 2003]{woosley:1988c}.  Since the models of
\citet{scheck:2007a} included only the central part of the star out to
the middle of the C/O shell at $r = 1.7\times 10^9\,$cm, the envelope
of the progenitor had to be added for the long-time simulations
presented here.

The $15.5\,M_\odot$ blue supergiant stellar progenitor model was
evolved by \citet{woosley:1988c} using a nuclear reaction network
including a large set of nuclei. However, \citet{scheck:2007a} did not
consider nuclear burning, and he included only a small number of
nuclear species ($p$, $n$, $\alpha$, and $^{54}\rm Mn$ representing
the heavy element fraction) in their simulations to save computational
costs.  For our initial model we kept the nuclear composition of the
3D explosion model of \citet{scheck:2007a} inside the shock radius,
redefining $^{54}\rm Mn$ into $^{56}\rm Ni$. The nuclear composition
at radii greater than the shock radius was taken from the
$15.5\,M_\odot$ progenitor model.

\citet{scheck:2007a} simulated one 3D hydrodynamic explosion model
with $2^\circ$ angular resolution, which reached a final time of
0.58\,s after bounce, and another one with $3^\circ$ angular
resolution, which covered an evolution time of 1\,s after core bounce.
Although both models were exploited for our study, we report here only
on simulations using the former model that according to the data of
\citet{scheck:2007a} has an (unsaturated) explosion energy of $0.6
\times 10^{51}\,$erg. The deformation of the shock front at $t \sim
0.5\,$s after core bounce and the asymmetry of the entropy
distribution behind the shock in the form of rising, inflating hot
plumes and infalling, narrow, cool downflows are shown in
Fig.~\ref{fig:inimod}. The structure of the explosion at this time
looks very similar to comparable results by
\citet{fryer:2002a}. Besides following the shock's propagation through
the stellar envelope of this initial model with our 3D code, we also
performed a 3D simulation where we artificially increased the
explosion energy of this initial model to $1.0\times 10^{51}\,$erg by
extending the thermal energy input as accomplishable by ongoing
neutrino heating.

\subsection{Equation of state}
\label{sec:eos}
We used a tabulated EOS \citep{timmes:2000a} that considers
contributions of an arbitrarily degenerate and relativistic
electron-positron gas, of a photon gas, and of a set of ideal
Boltzmann gases, consisting of the eight nuclear species ($n$, $p$,
$\alpha$, $^{12}\rm C$, $^{16}\rm O$, $^{20}\rm Ne$, $^{24}\rm Mg$,
and $^{56}\rm Ni$) included in our initial model. Note that the
pre-shock layers contained only a small region of silicon at the 
time the explosion model of \citet{scheck:2007a} was adopted as initial
data for our long-time supernova runs. Since nuclear burning was not
taken into account in our simulations, but it is possible that this
silicon will be explosively burned to nickel, we lumped the small 
amount of silicon and the iron-group
elements in the postshock layer together to what we 
followed as $^{56}\rm Ni$ in our hydrodynamic calculations.

\subsection{Simulation runs}
\label{sec:simulations}
We performed a set of thirteen 2D and three 3D simulations and compared 
the results obtained in three dimensions to those of the corresponding
2D models. In this paper we focus our discussion on one 3D model that
has an explosion energy of $10^{51}\,$ergs. The main findings for this
particular 3D model are qualitatively similar to those we obtained for
less energetic 3D explosions, but the resulting maximum values for the
asymptotic clump velocities vary with the explosion energy roughly as
$v_{\mathrm{clump}}\propto \sqrt{E_\mathrm{exp}}$.

The 2D simulations reported here were started from different
meridional slices of the 3D initial explosion model. This ensures as
much similarity of the initial conditions as possible and the same
numerical resolution for the 2D and 3D runs. But the differences
between the chosen slices give rise to some variation of the initial
asymmetry and explosion energy among the various 2D models, which
manifests itself both in different clump structures and velocities,
and the amount of mixing in the stellar envelope.

A set of 13 meridional 2D slices was placed at azimuthal angles 
of 4, 36, 76, 108, 128, 148, 180, 220, 228, 252, 292, 324, and 360 
degrees. Three of these slices (at 128, 220, and 228 degrees) were
chosen at locations where the growth of 
Rayleigh-Taylor fingers was observed to be particularly strong in 3D.
This guaranteed that such regions were not missed by our
comparison of 2D and 3D calculations.
It turned out that the largest 3D Rayleigh-Taylor structures at 
late stages develop
in the directions of the biggest initial buoyant bubbles, expanding
towards the bottom edge and upper right corner (at six o'clock and 
one o'clock, respectively) of Fig.~\ref{fig:inimod}. 

Both in the 3D and 2D simulations we neglected the influence of
gravity on the motion of the ejecta. While this has no important
impact on the dynamics of the expanding ejecta, the amount of fallback
is underestimated. However, that way we could avoid the accumulation
of mass near the inner (reflecting) radial boundary which would have
implied a considerably more restrictive CFL condition.

%%%%%%%%%%%%%%%%%%%%%%%%%%%%%%%%%%%%%%%%%%%%%%%%%%%%%%%%
%% FIGURE 2
%%%%%%%%%%%%%%%%%%%%%%%%%%%%%%%%%%%%%%%%%%%%%%%%%%%%%%%%
\begin{figure*}
\begin{center}
\plottwo{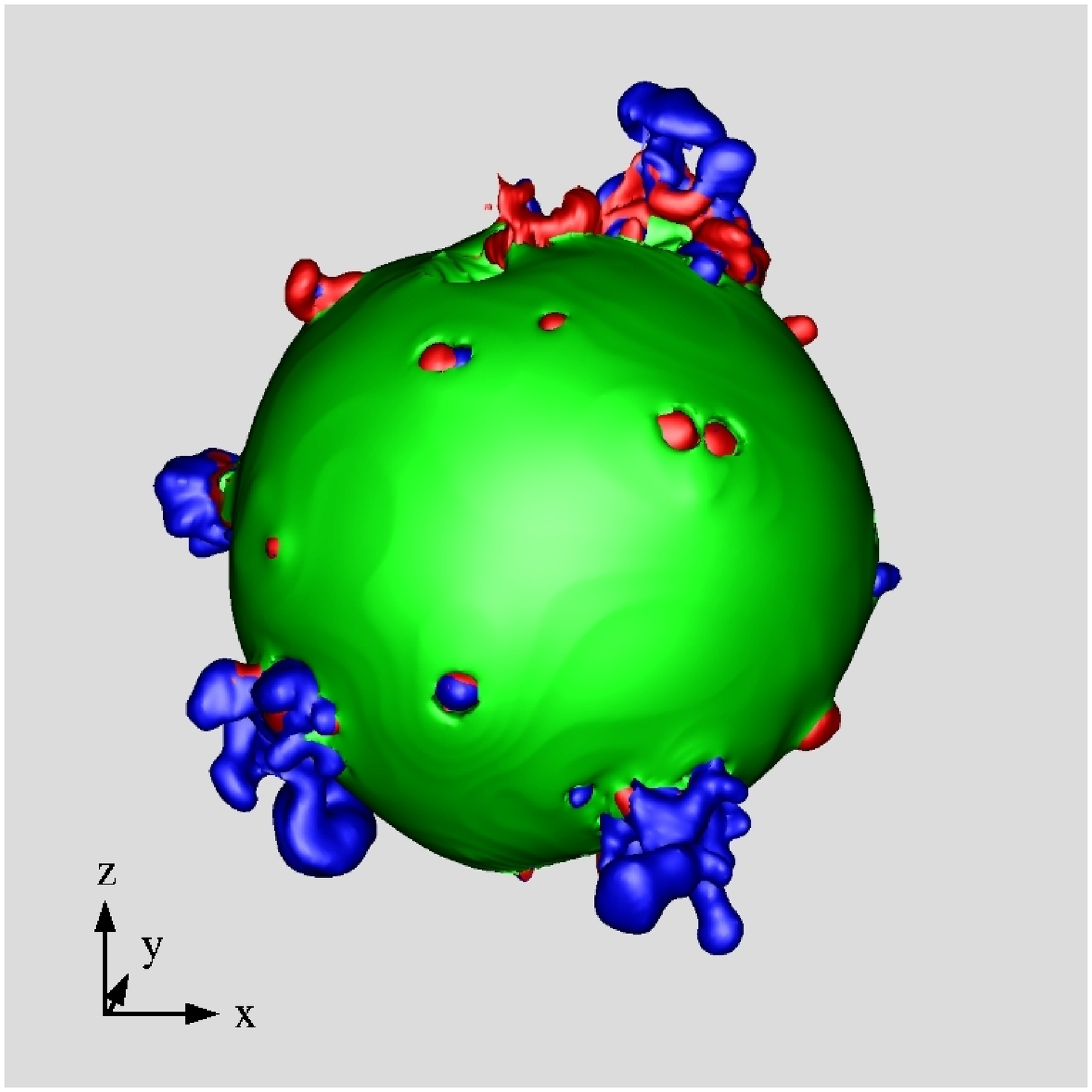}{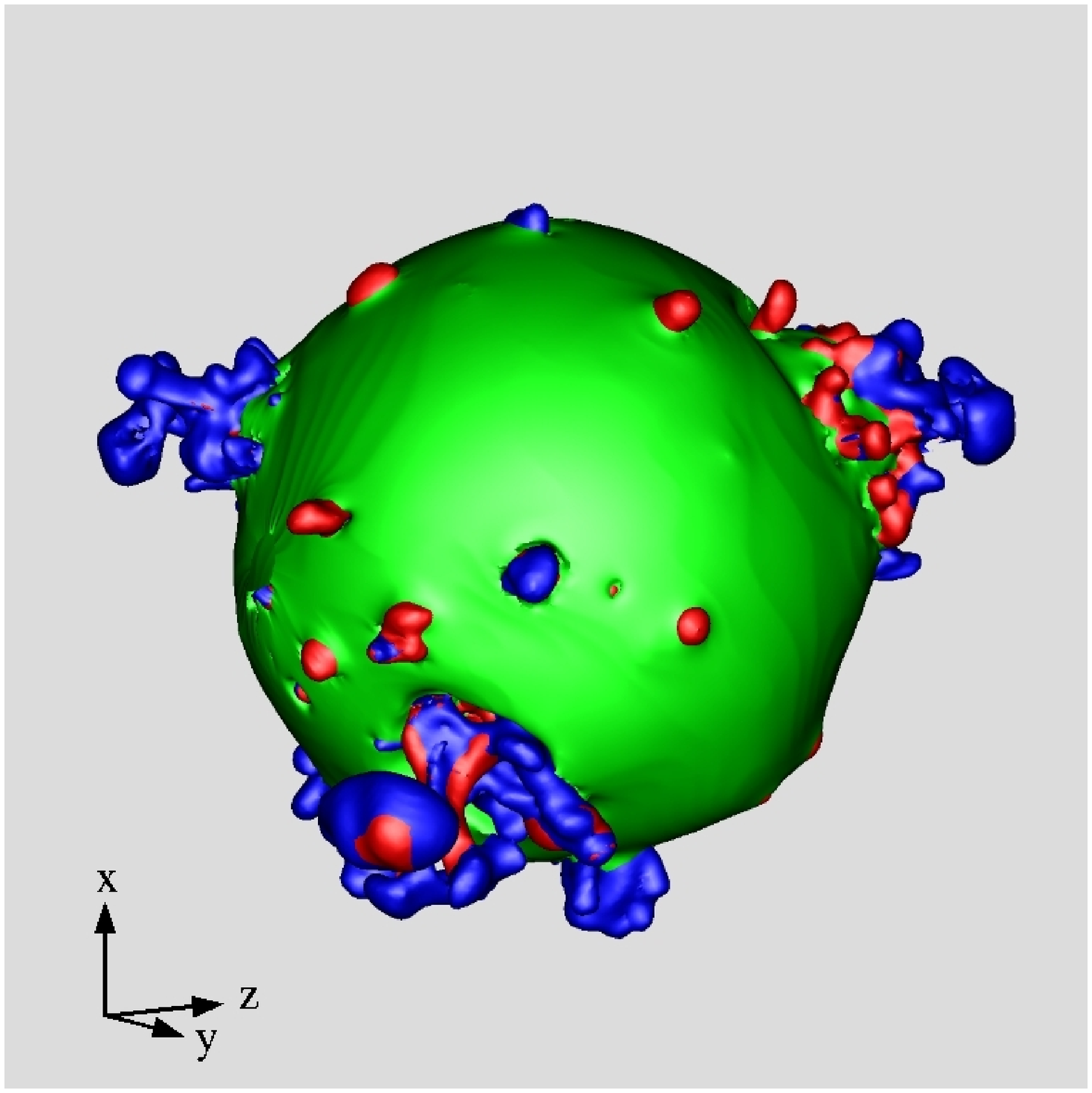}\\[2mm]
\plottwo{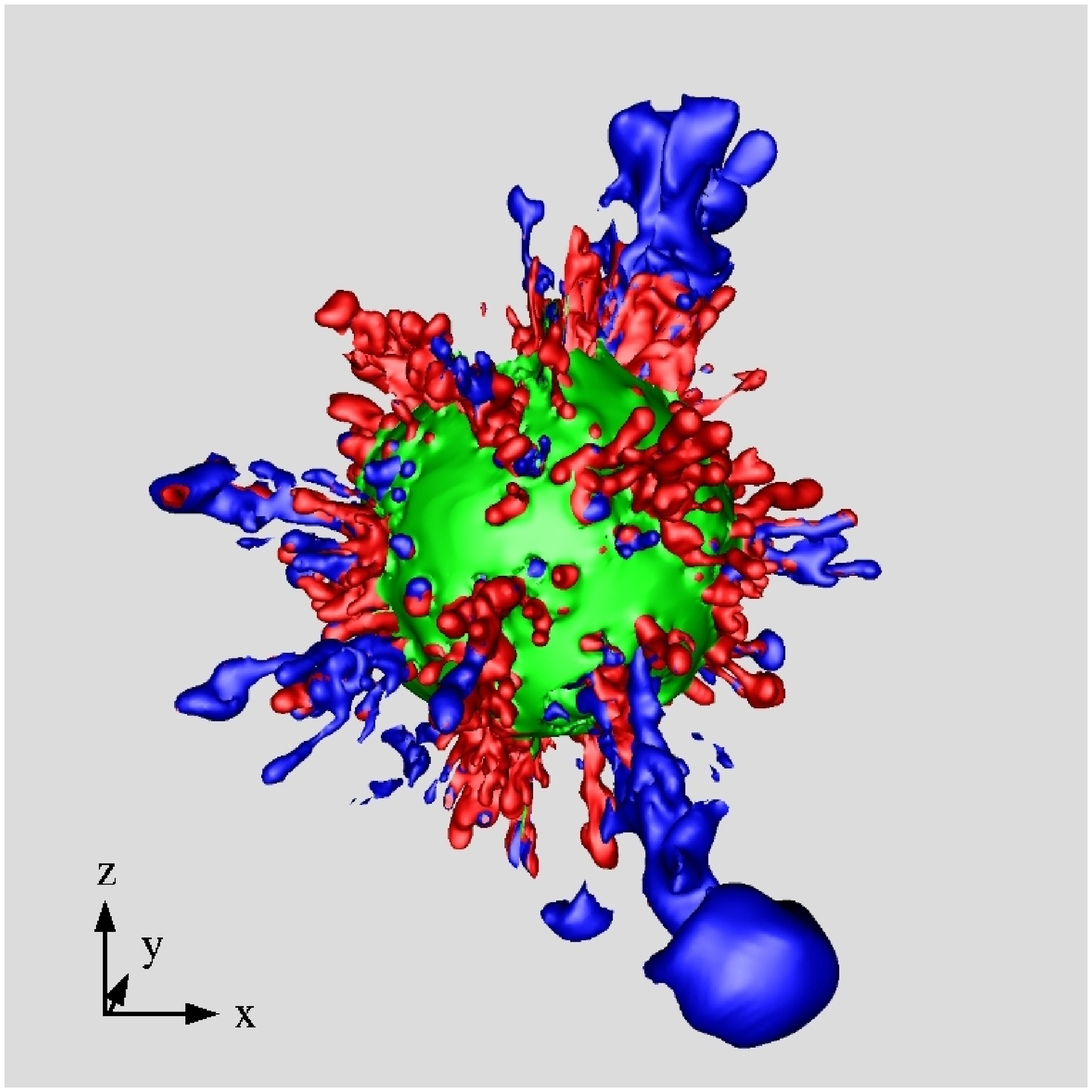}{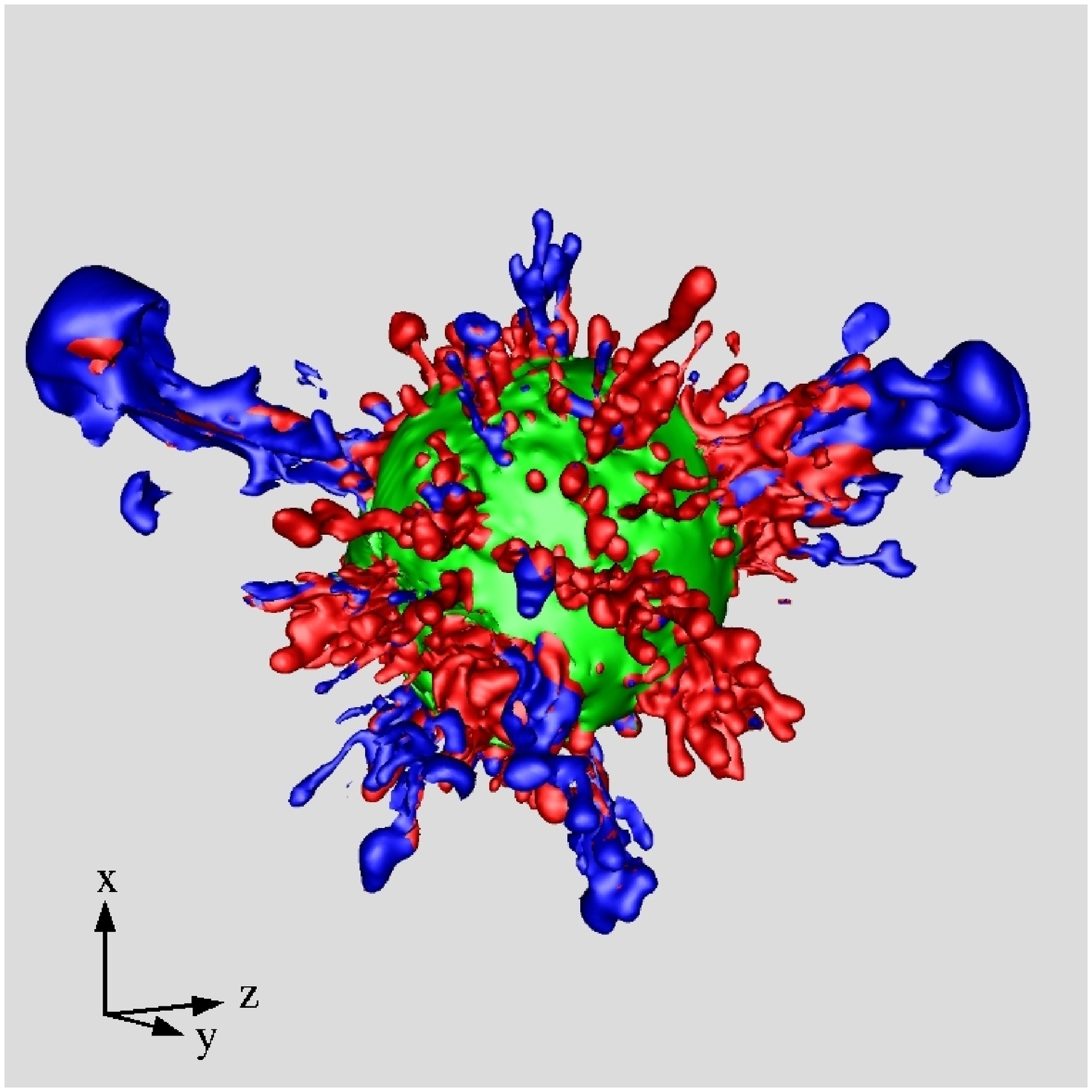}
\caption{\label{fig:3Dsurfaces}
  Surfaces of the radially outermost locations with constant mass
  fractions of $\sim 3$\% for carbon (green), and oxygen (red), and
  of $\sim 7$\% for nickel (blue).  The upper two panels show the
  directional asymmetries from two different viewing directions at
  350$\,$s after core bounce when the metal clumps begin to enter
  the helium layer of the star. The lower two panels display the 
  hydrodynamic instabilities at about 9000$\,$s shortly after the
  supernova shock has broken out of the stellar surface.
  The side length of the upper panels is about $5\times
  10^{11}\,$cm, of the lower panels $7.5\times 10^{12}\,$cm.}
\end{center}
\end{figure*}
%%%%%%%%%%%%%%%%%%%%%%%%%%%%%%%%%%%%%%%%%%%%%%%%%%%%%%%%

%%%%%%%%%%%%%%%%%%%%%%%%%%%%%%%%%%%%%%%%%%%%%%%%%%%%%%%%
%% FIGURE 3
%%%%%%%%%%%%%%%%%%%%%%%%%%%%%%%%%%%%%%%%%%%%%%%%%%%%%%%%
\begin{figure}
\begin{center}
\plotone{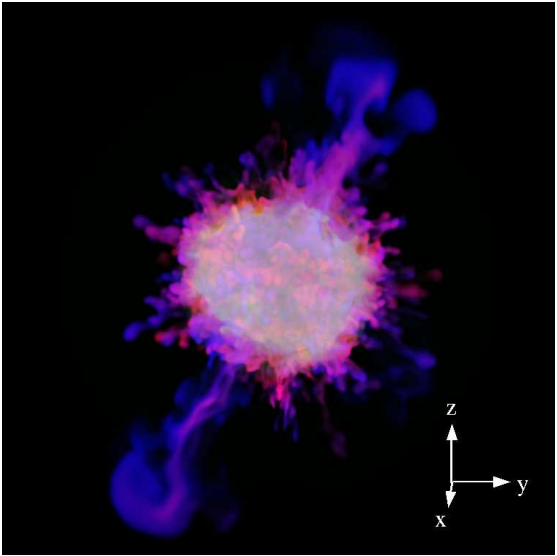}
\caption{\label{fig:3Dvolume}
  Volume rendered image of the element distribution at about
  9000$\,$s (corresponding to the lower two panels in 
  Fig.~\ref{fig:3Dsurfaces}, but displayed for yet 
  another viewing angle). The blue structures contain a 
  dominant mass fraction of nickel, the red fingers mostly 
  oxygen, and green is associated with carbon. A mixture of 
  nickel and oxygen appears in pink. Note that the green 
  visible in Fig.~\ref{fig:3Dsurfaces} 
  has been subsumed into the white glow
  due to the contamination with other colors as a consequence of
  the volume rendering and not of true mixing. The side length 
  of the displayed volume is about $7.5\times 10^{12}\,$cm.
  It is remarkable that the two long Rayleigh-Taylor fingers 
  from this perspective create the impression of a jet and an 
  anti-jet, although no jet-creating mechanism was in action
  at the center of the explosion and the long Rayleigh-Taylor
  structures do actually not point into exactly opposite (i.e., 
  antiparallel) directions (cf.\ Fig.~\ref{fig:3Dsurfaces}).
}
\end{center}
\end{figure}
%%%%%%%%%%%%%%%%%%%%%%%%%%%%%%%%%%%%%%%%%%%%%%%%%%%%%%%%

%%%%%%%%%%%%%%%%%%%%%%%%%%%%%%%%%%%%%%%%%%%%%%%%%%%%%%%%
%% FIGURE 4
%%%%%%%%%%%%%%%%%%%%%%%%%%%%%%%%%%%%%%%%%%%%%%%%%%%%%%%%
\begin{figure*}
\begin{center}
\plottwo{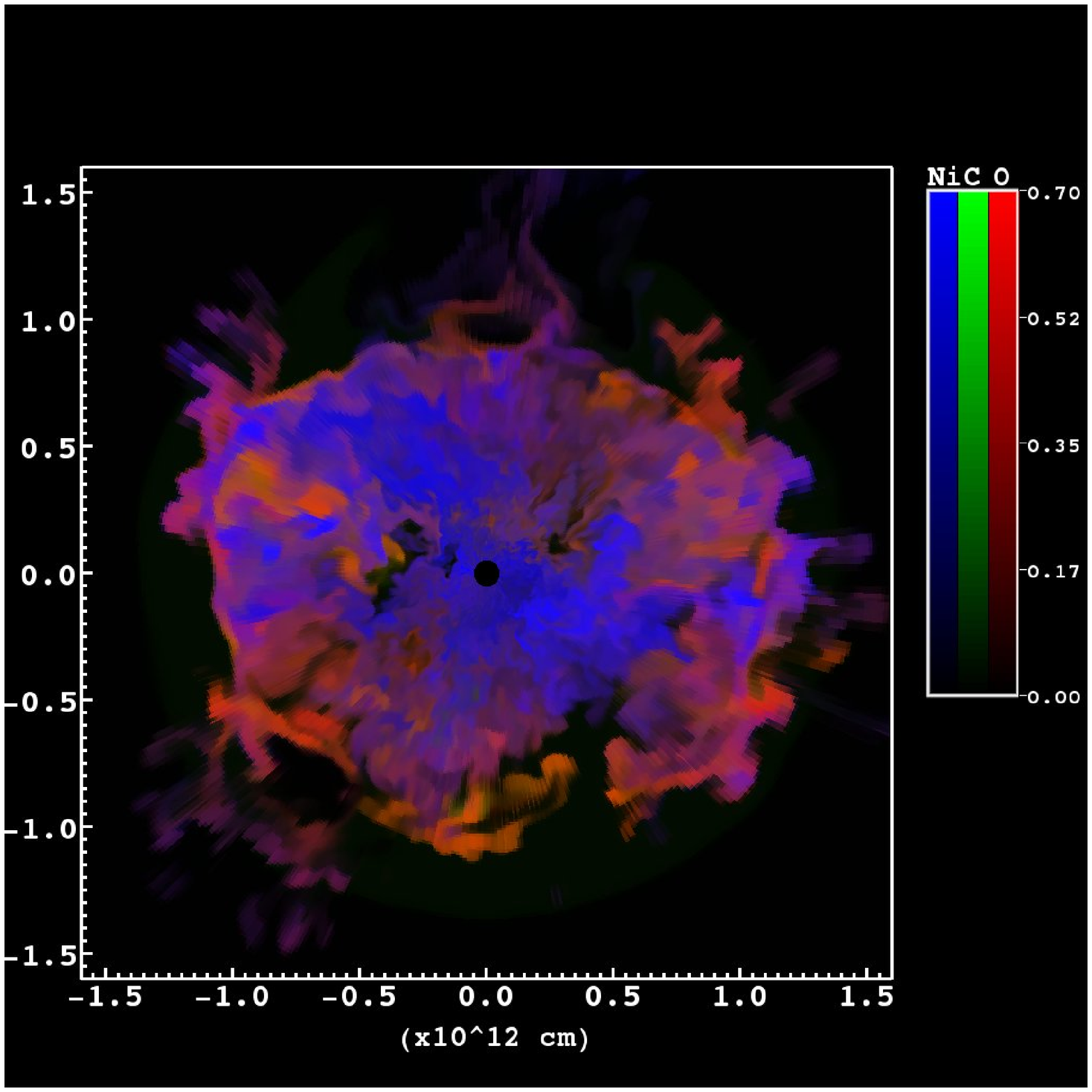}{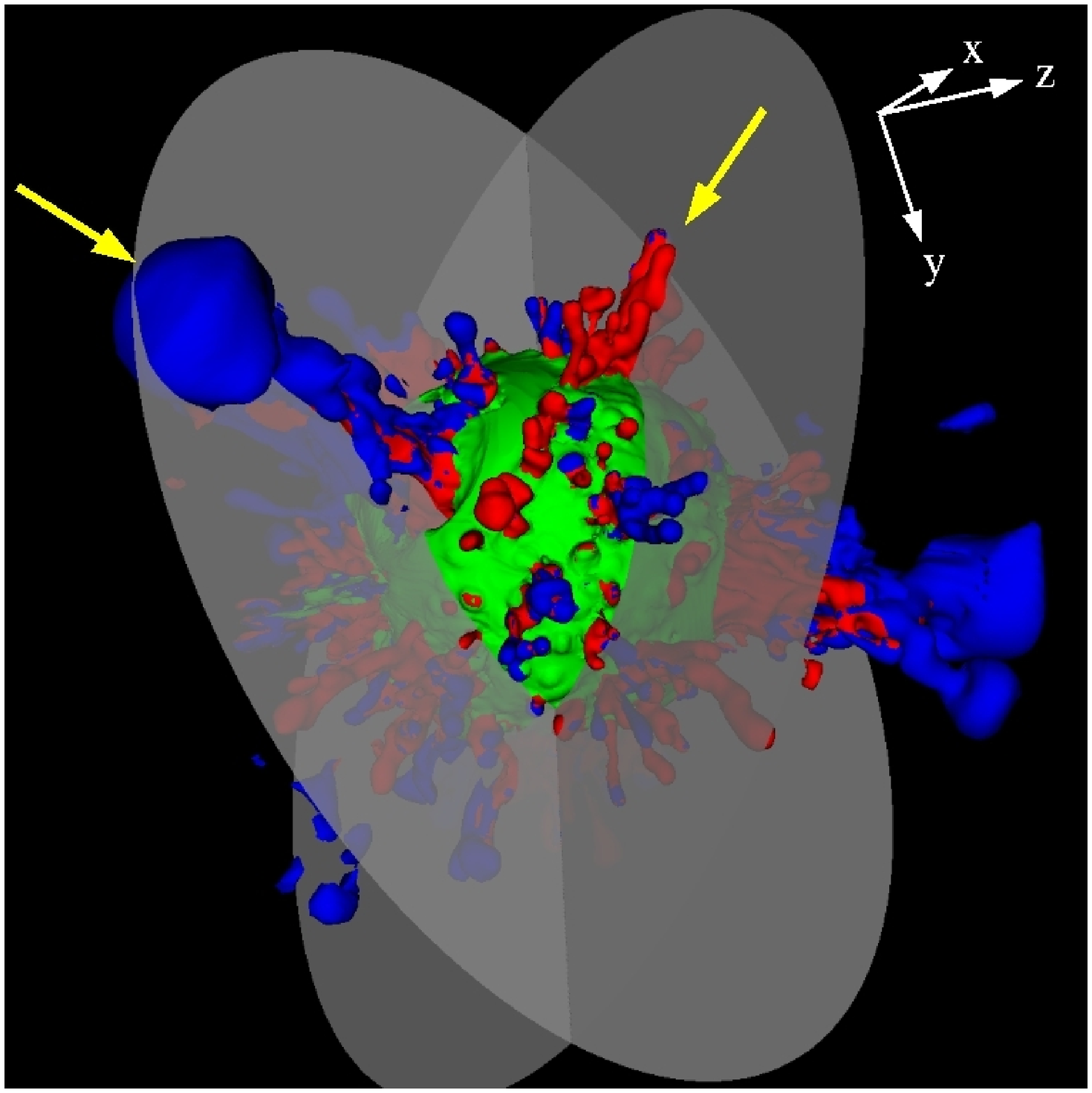}\\[2mm]
\plottwo{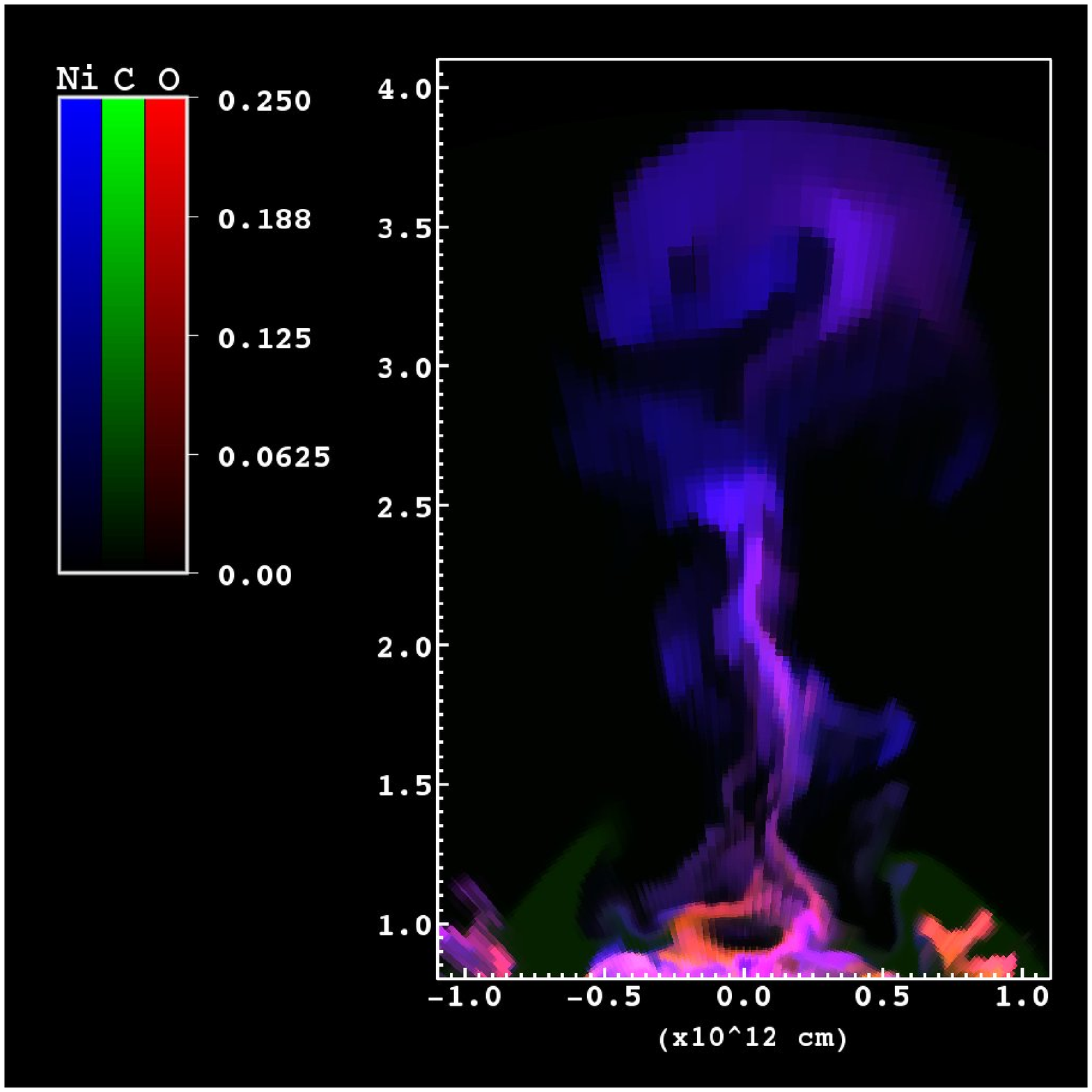}{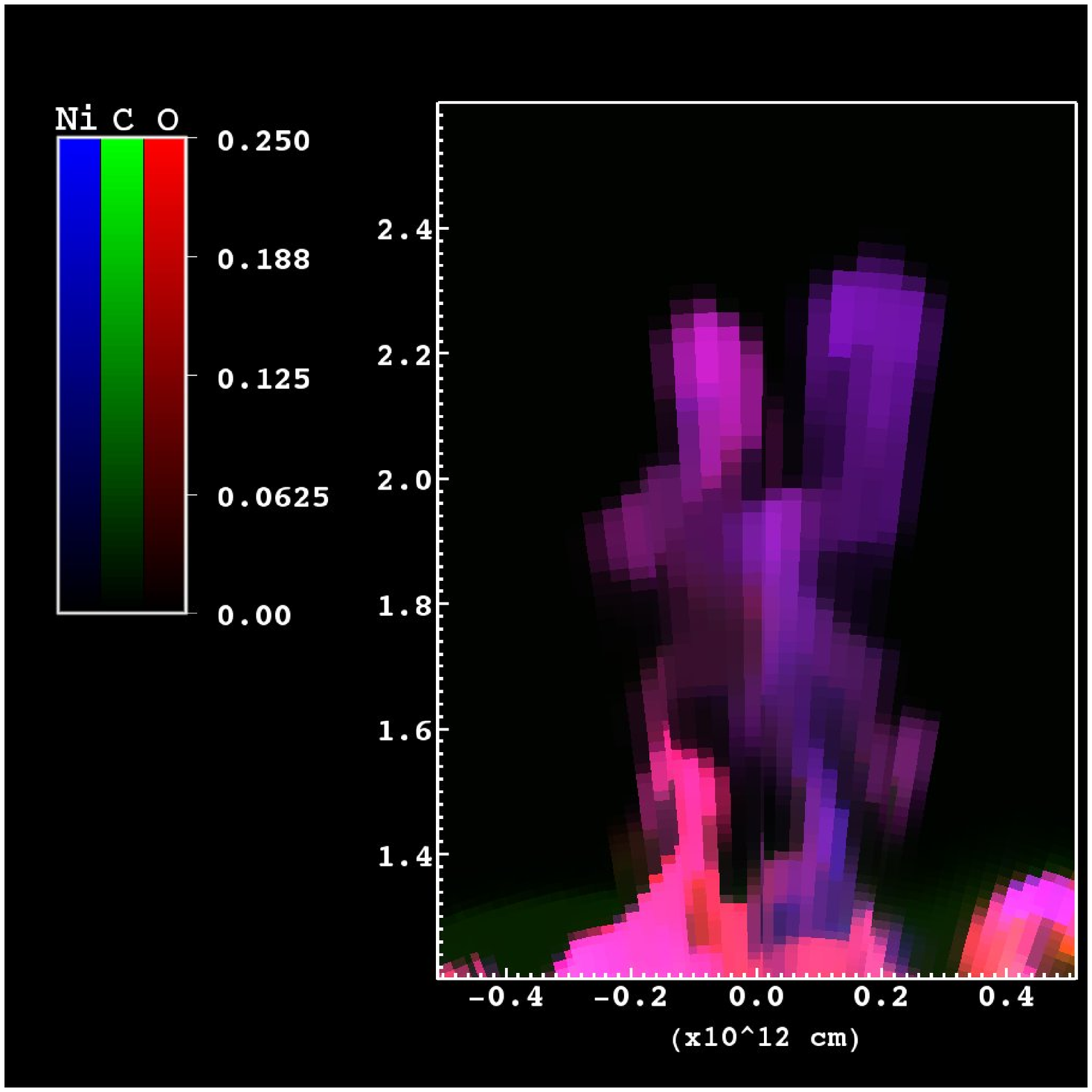}
\caption{\label{fig:3Dcuts}
  Composition information on 2D cuts through the 3D model structure.
  Blue corresponds to nickel, red to oxygen, green to carbon.
  Note that for better visibility of the dynamical variation, 
  different color ranges were chosen for the mass fractions in 
  the global-view image ({\em top left}) and in the local zooms 
  ({\em bottom}).
  {\em Upper left:} Mass fractions of Ni, O, and C 
  color coded in a cross sectional plane through the mixed core.
  The cut plane is indicated in the {\em upper right} panel and
  crosses the big mushroom that is enlarged in the {\em lower left}
  panel (this feature grows out of the core at the 12:00 o'clock 
  position),
  but the basic features of the core structure do not depend much
  on the chosen location of the plane through the grid origin.
  {\em Upper right:} Cutting planes through two representative 
  Rayleigh-Taylor structures, the big mushroom growing towards 
  10 o'clock, and the multi-finger feature extending towards
  12:30 o'clock. Both features are marked by thick yellow arrows.
  {\em Lower left:} Composition in the cut plane through the 
  big Rayleigh-Taylor mushroom.
  {\em Lower right:} Composition in the cut plane through the
  multi-finger region.
  All extended plumes and fingers carry a mix of heavy elements
  with different mass fractions and relative abundances, but
  also large amounts of hydrogen and helium (between $\sim$50 
  and $\sim$70\% of the mass with considerable spatial 
  variations inside individual structures). }
\end{center}
\end{figure*}
%%%%%%%%%%%%%%%%%%%%%%%%%%%%%%%%%%%%%%%%%%%%%%%%%%%%%%%%

\section{Results}

\subsection{Dynamical evolution}
\label{sec:dynevol}
The dynamical evolution of the explosion after its launch by neutrino
energy deposition and the growth of instabilities in the considered
15.5$\,M_\odot$ progenitor were described in much detail by
\citet{kifonidis:2003a}. It was shown there that the asymmetries
associated with the convective activity developing in the postshock
region during the neutrino-heating phase act as seeds of secondary
Rayleigh-Taylor instabilities at the composition interfaces of the
exploding star. At about 100$\,$s dense Rayleigh-Taylor fingers
containing the metals (C, O, Si, iron-group elements) have grown out
of a compressed shell of matter left behind by the shock passing
through the Si/O and (C+O)/He interfaces. These fingers grow quickly
in length and while extending into the helium shell, fragment into
ballistically moving clumps and filaments that propagate faster than
the expansion of their environment.

While for a 2D model with explosion energy around 1.8$\times
10^{51}\,$erg the Si and Ni containing structures still move with
nearly 4000$\,$km$\,$s$^{-1}$ (oxygen has velocities up to even
5000$\,$km$\,$s$^{-1}$) at 300$\,$s, a strong deceleration occurs when
the metal clumps enter the relatively dense He-shell that forms after
the shock passage through the He/H interface.  At $t\ga 10,000\,$s the
metal carrying clumps have dissipated their excess kinetic energy and
propagate with the same speed as the helium material in their
surroundings. In the presence of a hydrogen envelope and the
corresponding deceleration of the shock as it propagates through the
inner regions of the hydrogen layer (in which case the helium ``wall''
below the He/H interface builds up), \citet{kifonidis:2003a} could not
observe any metal clumps that achieve to penetrate into the hydrogen
envelope, in contrast to what was observed in the case of SN\,1987A.

The 2D calculations performed in the course of the present work
confirm these findings (when the velocities are appropriately rescaled
with $\sqrt{E_\mathrm{exp}}$ to account for the lower explosion
energies of $E_{\mathrm{exp}}\la 10^{51}\,$erg considered here
compared to roughly twice this value employed by
Kifonidis et al.\ 2003).  The evolution as well as the final results
for the mass distributions of different chemical elements in velocity
space and mass space are in good quantitative agreement with the
models of \citet{kifonidis:2003a}, although the latter had much
superior numerical resolution because of the use of a sophisticated
adaptive mesh refinement algorithm. We are therefore confident that at
least with respect to gross (quasi-1D) features, the results presented 
here are numerically converged, even if many of the structural
details, including plumes and bullets, may not be numerically 
converged. 

\citet{kifonidis:2006a}, investigating explosions also with
$E_{\mathrm{exp}}\sim 2\times 10^{51}\,$erg, proposed a cure of the
metal-H mixing problem by invoking a sufficiently large dipolar or
quadrupolar asymmetry of the beginning explosion with a globally
aspherical shock wave.  This was motivated by recent hydrodynamical
studies that show that large-amplitude, low-$\ell$ mode oscillations
($\ell$ defining the order of an expansion in spherical harmonics) due
to the standing accretion shock instability (SASI) can be an important
ingredient in the explosion mechanism and in the sequence of processes
that leads to observed asymmetries of supernova explosions and neutron
star kicks \citep[see e.g.,][]{blondin:2003a, blondin:2007a,
  scheck:2004a, scheck:2006a, scheck:2007a, scheck:2008a,
  burrows:2006a, burrows:2007a, marek:2009a}. This on the one hand led
to higher initial maximum velocities of the metal-rich clumps and thus
their faster propagation through the He core on a timescale shorter
than the build-up of the thick He ``wall'' that prevented their
penetration into the hydrogen shell in the older calculations.  On the
other hand it triggered the growth of Richtmyer-Meshkov instability
(RMI) at the He/H interface due to the deposition of vorticity by the
shock. This led to strong mixing of hydrogen inward in mass space and
down to low velocities, an effect that has turned out to be necessary
to explain the shape of the light curve maximum in the case of
SN\,1987A \citep{utrobin:2004a}.

These conclusions, however, were based on 2D models. Therefore the
question remained whether 3D might yield differences and whether also
in 3D one would have to advocate a pronounced 
global low-$\ell$ mode asphericity
of the beginning explosion as an explanation of the high velocities
($\ga 3000\,$km$\,$s$^{-1}$) of nickel clumps in the hydrogen shell
and of the strong inward mixing of hydrogen observed in SN\,1987A. The
3D simulations performed in the present work demonstrate that this is not
the case and the mentioned observational features can be accounted for
by 3D models even with explosion energies around $10^{51}\,$ergs for
the considered progenitor star.

In order to quantify the global shock deformation of our 3D initial
models compared to the 2D cases investigated by \citet{kifonidis:2006a},
we did not perform an analysis in terms of an expansion in spherical
harmonics modes. Instead, sizeable differences become obvious already
when a tri-axial ellipsoid is constructed as 
envelope of the bumpy shock surface obtained by \citet{scheck:2007a}
such that the major axis of the ellipsoid coincides with the direction
of the strongest shock expansion. While the 3D model used for the 
calculations of the present paper has an axis ratio of 1.04:1.02:1.00
at $t = 0.58\,$s after bounce --- another 3D explosion run of
\citet{scheck:2007a} yields 1.16:1.06:1.00 at $t = 1\,$s ---, the shocks 
of the 2D configurations studied by \citet{kifonidis:2006a} typically
had much larger axis ratios of about 1.5:1.0.

The deformed shock causes strong Richtmyer-Meshkov instability (RMI)
to develop at the (C+O)/He and He/H interfaces in the 2D models
discussed here. Compared to the runs of \citet{kifonidis:2006a}, 
however, the smaller asphericity of the shock leads to a much slower
growth of the instability. In our 2D runs the corresponding vortices
at the He/H boundary at 13000$\,$s have reached a size relative to
the average interface radius similar to what
was observed at 3000$\,$s in the models of \citet{kifonidis:2006a}.
We note in passing that apart from resolution-dependent fine
structure, the results of the latter paper could be very well
reproduced concerning the RMI vortex shape, size, and locations
by simulations performed for the same initial conditions but with the
grid setup and resolution used in the present work 
(see \citet{hammer:2009}). The growth of the RMI observed
in average cases of our 2D calculations with only weak
low-$\ell$ mode shock deformation --- in agreement with analytic
growth rate estimates; \citet{hammer:2009} --- is much too slow 
to produce any significant extent of hydrogen-helium mixing so that
the final outcome of our 2D models basically confirms the findings 
of \citet{kifonidis:2003a}\footnote{However, there is a considerable
spread of the 2D results depending on the choice of the meridional
plane of the 2D slice and the variation of the conditions between the
different planes. Slices with somewhat larger initial shock deformation
show stronger late mixing into the hydrogen shell than the more typical 
``average'' 2D slices (see Fig.~\ref{fig:HNivsvel}).}.
In the 3D models RMI distortions can be
seen at the (C+O)/He interface and are likely to contribute to the
turbulent mixing of the metal core with the helium shell
of the exploding star. At the H/He interface, however, where the
shock is very close to being spherical, no clear RMI activity
becomes visible before it is penetrated by fast, metal-carrying
clumps that have been able to pass through the 
helium layer with still high velocities.

%%%%%%%%%%%%%%%%%%%%%%%%%%%%%%%%%%%%%%%%%%%%%%%%%%%%%%%%
%% FIGURE 5
%%%%%%%%%%%%%%%%%%%%%%%%%%%%%%%%%%%%%%%%%%%%%%%%%%%%%%%%
\begin{figure}
\begin{center}
\plotone{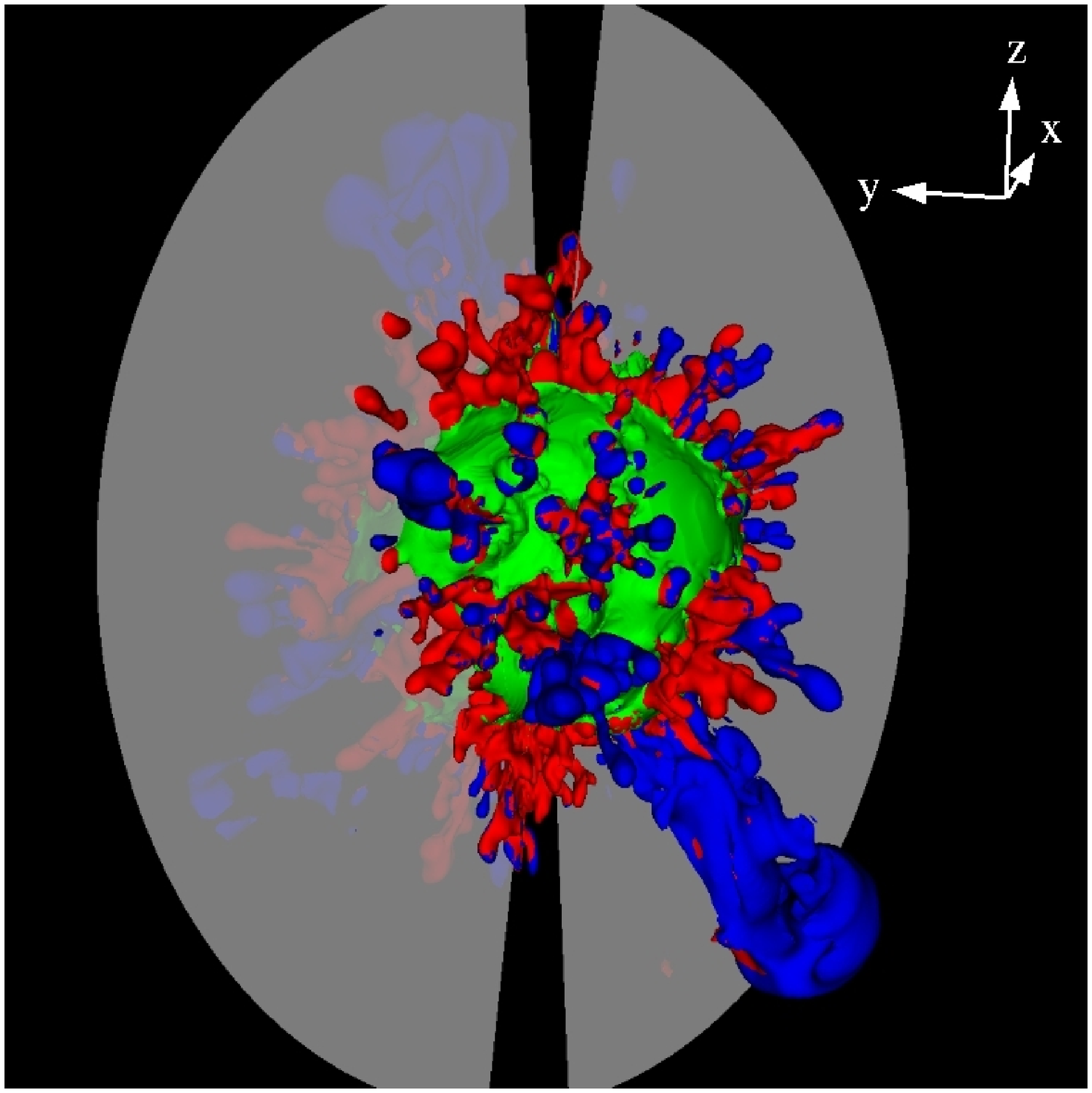}
\caption{\label{fig:2Dcut}
  Plane of the meridional cut at an azimuthal angle 
  of 148 degrees, which was the basis of the 2D simulation
  whose results are plotted in comparison to 3D in 
  Figs.~\ref{fig:3D2Dvsvel} and \ref{fig:3D2Dvsmass}.
  Note that we chose a ``typical'' (average)
  direction for the cut instead of an extreme one, while 
  the variation of the 2D results in different directions 
  is indicated in Fig.~\ref{fig:HNivsvel}.
}
\end{center}
\end{figure}
%%%%%%%%%%%%%%%%%%%%%%%%%%%%%%%%%%%%%%%%%%%%%%%%%%%%%%%%

%%%%%%%%%%%%%%%%%%%%%%%%%%%%%%%%%%%%%%%%%%%%%%%%%%%%%%%%
%% FIGURE 6
%%%%%%%%%%%%%%%%%%%%%%%%%%%%%%%%%%%%%%%%%%%%%%%%%%%%%%%%
\begin{figure*}
\begin{center}
\includegraphics[angle=90,width=.40\textwidth]{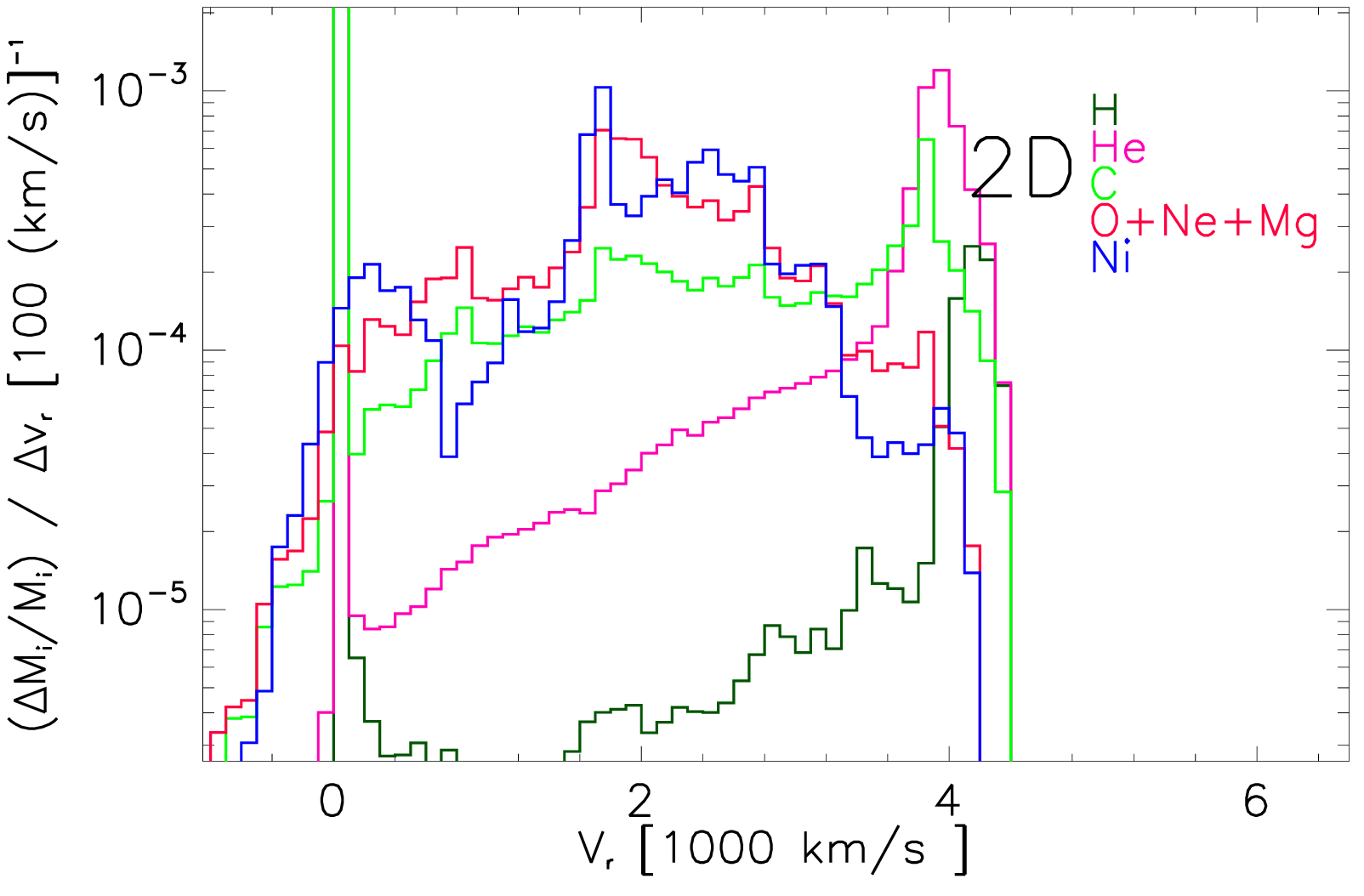}
\includegraphics[angle=90,width=.40\textwidth]{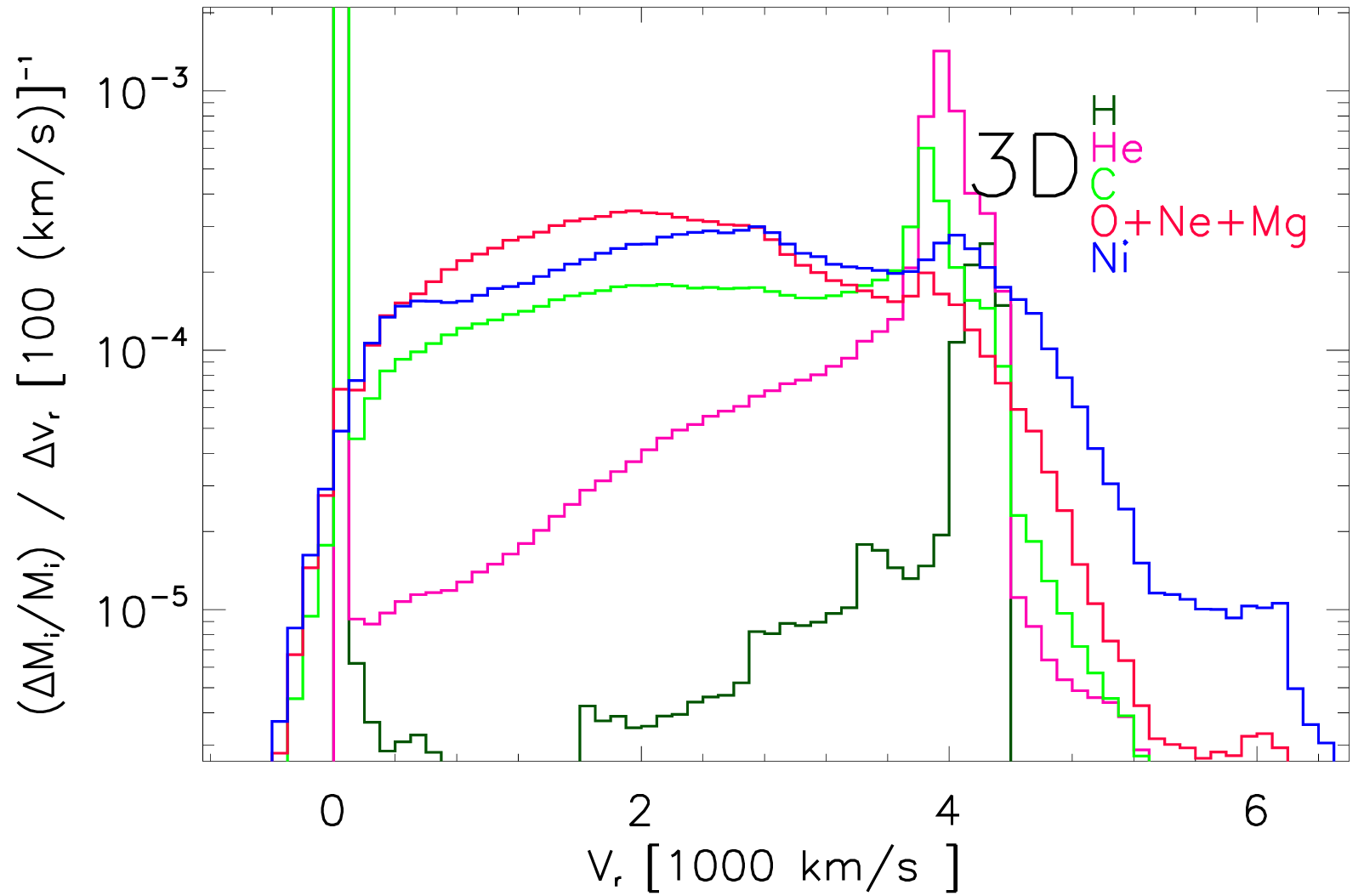}\\
\includegraphics[angle=90,width=.40\textwidth]{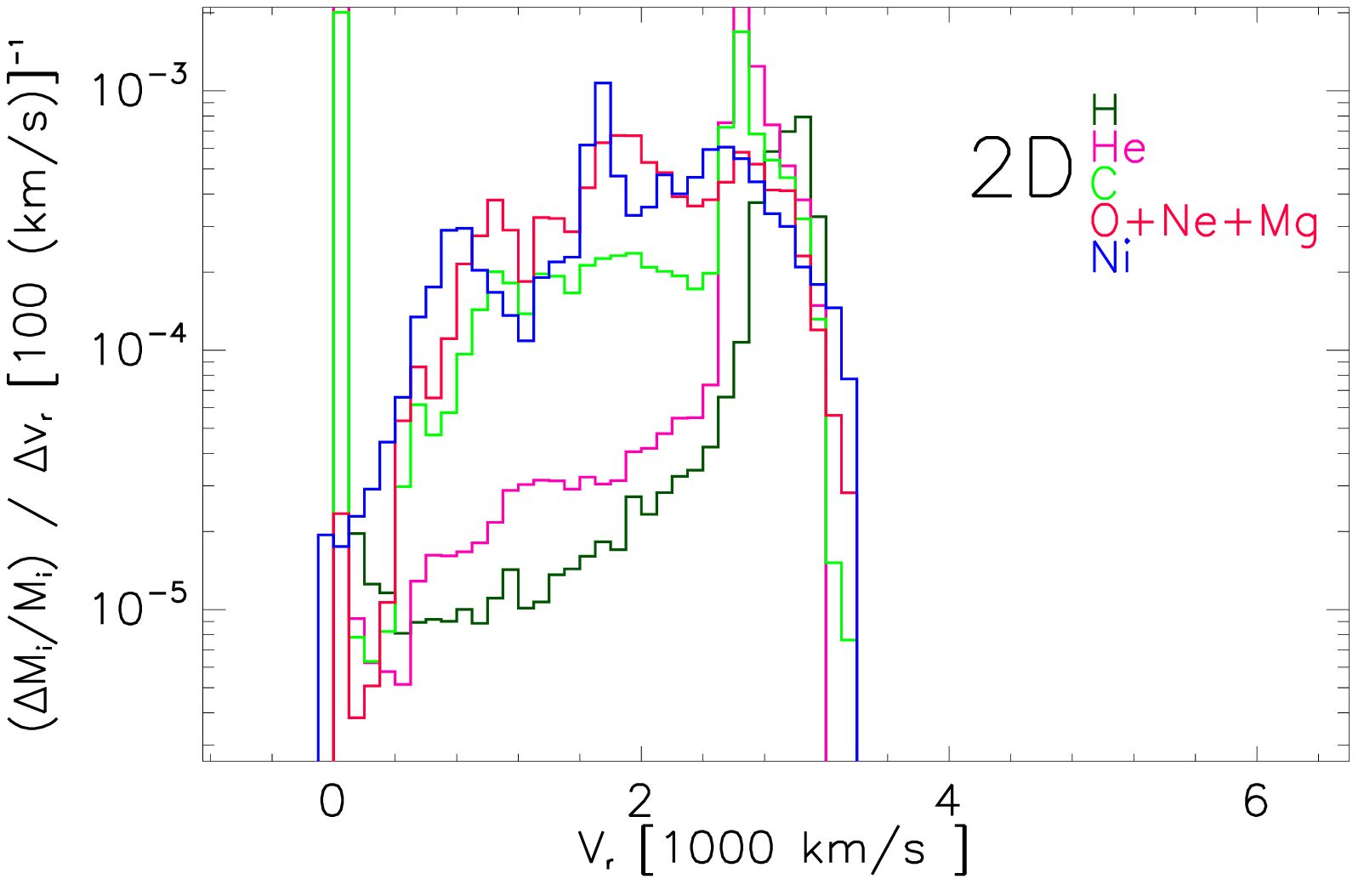}
\includegraphics[angle=90,width=.40\textwidth]{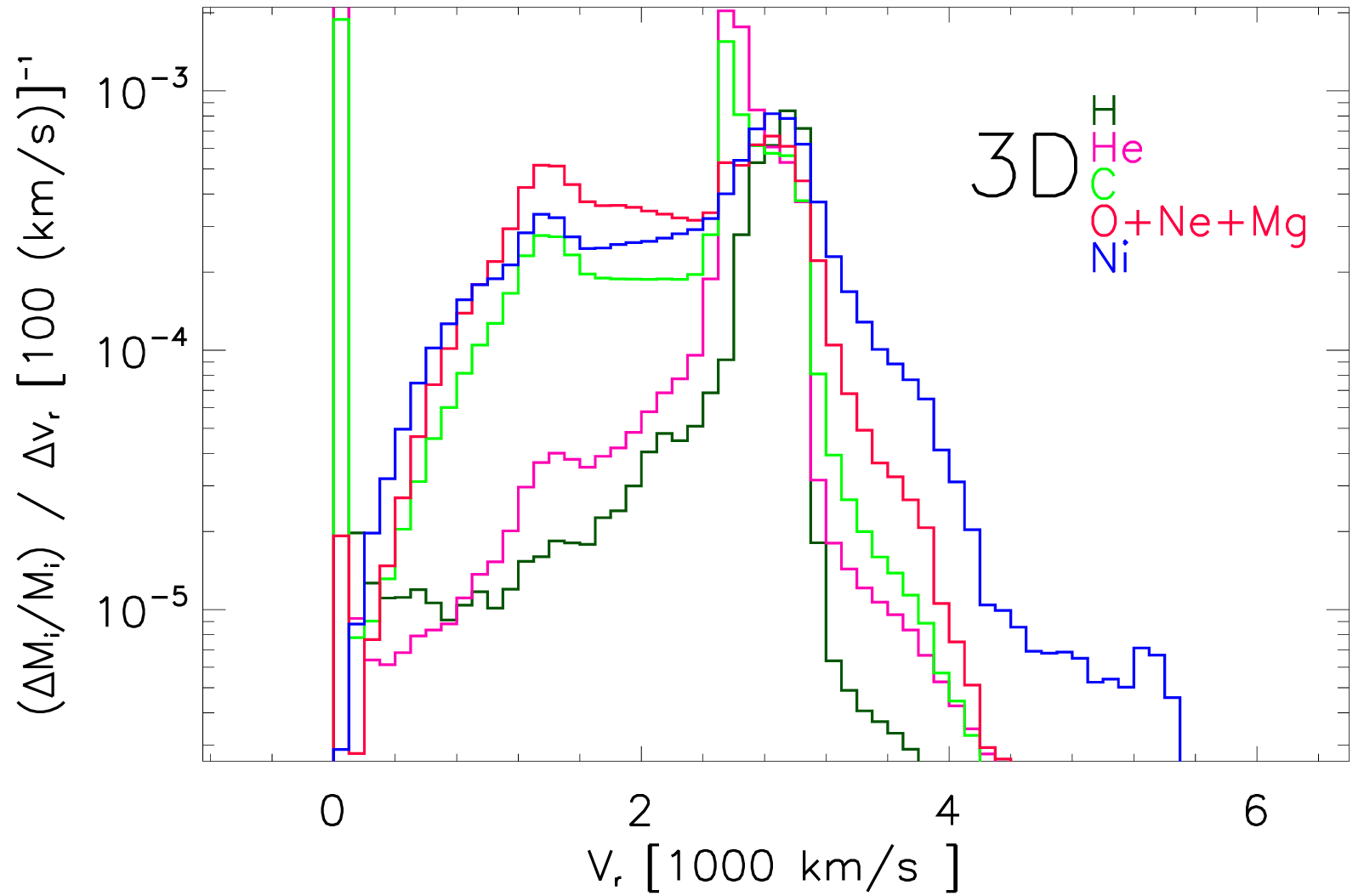}\\
\includegraphics[angle=90,width=.40\textwidth]{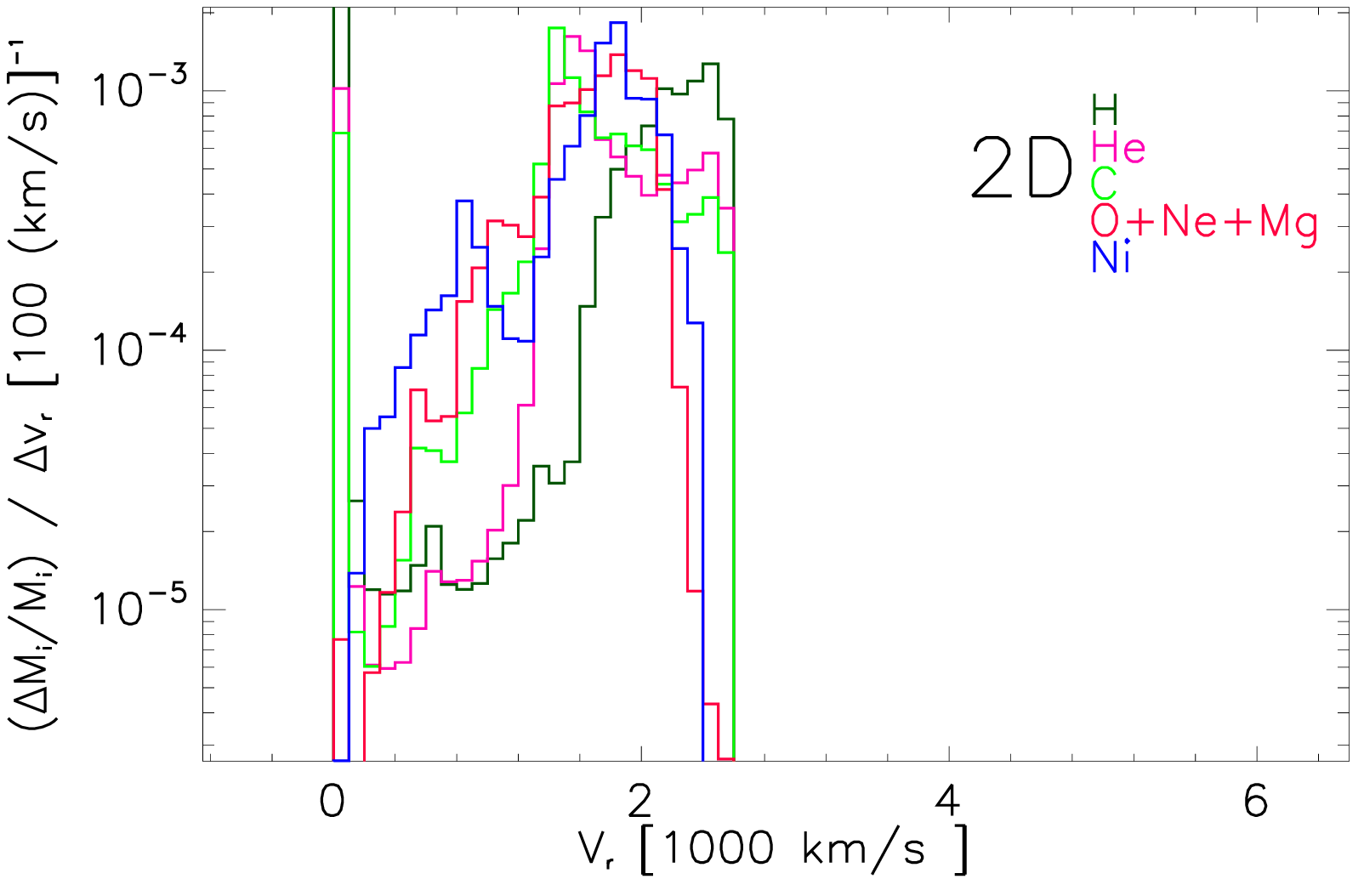}
\includegraphics[angle=90,width=.40\textwidth]{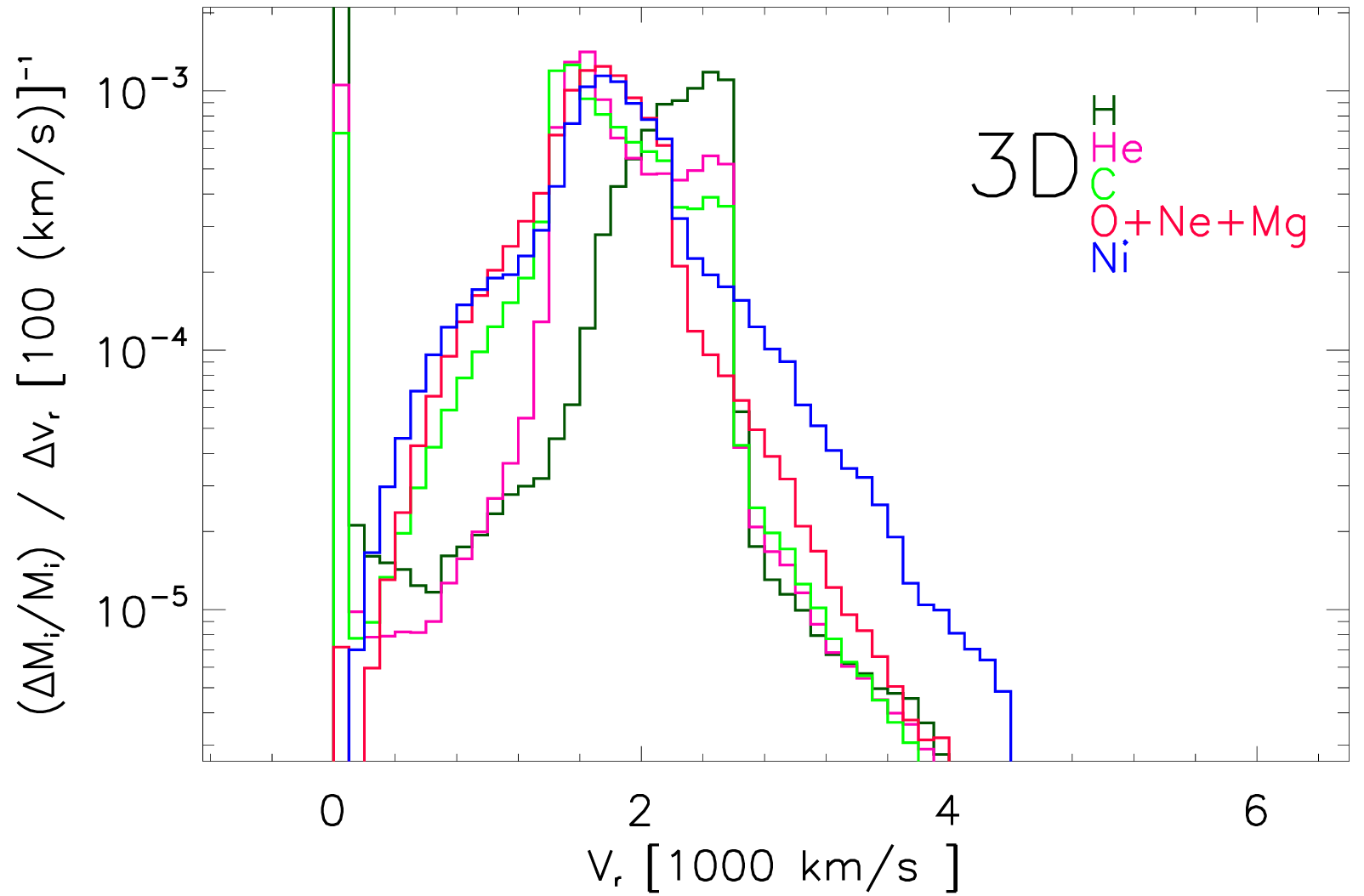}\\
\includegraphics[angle=90,width=.40\textwidth]{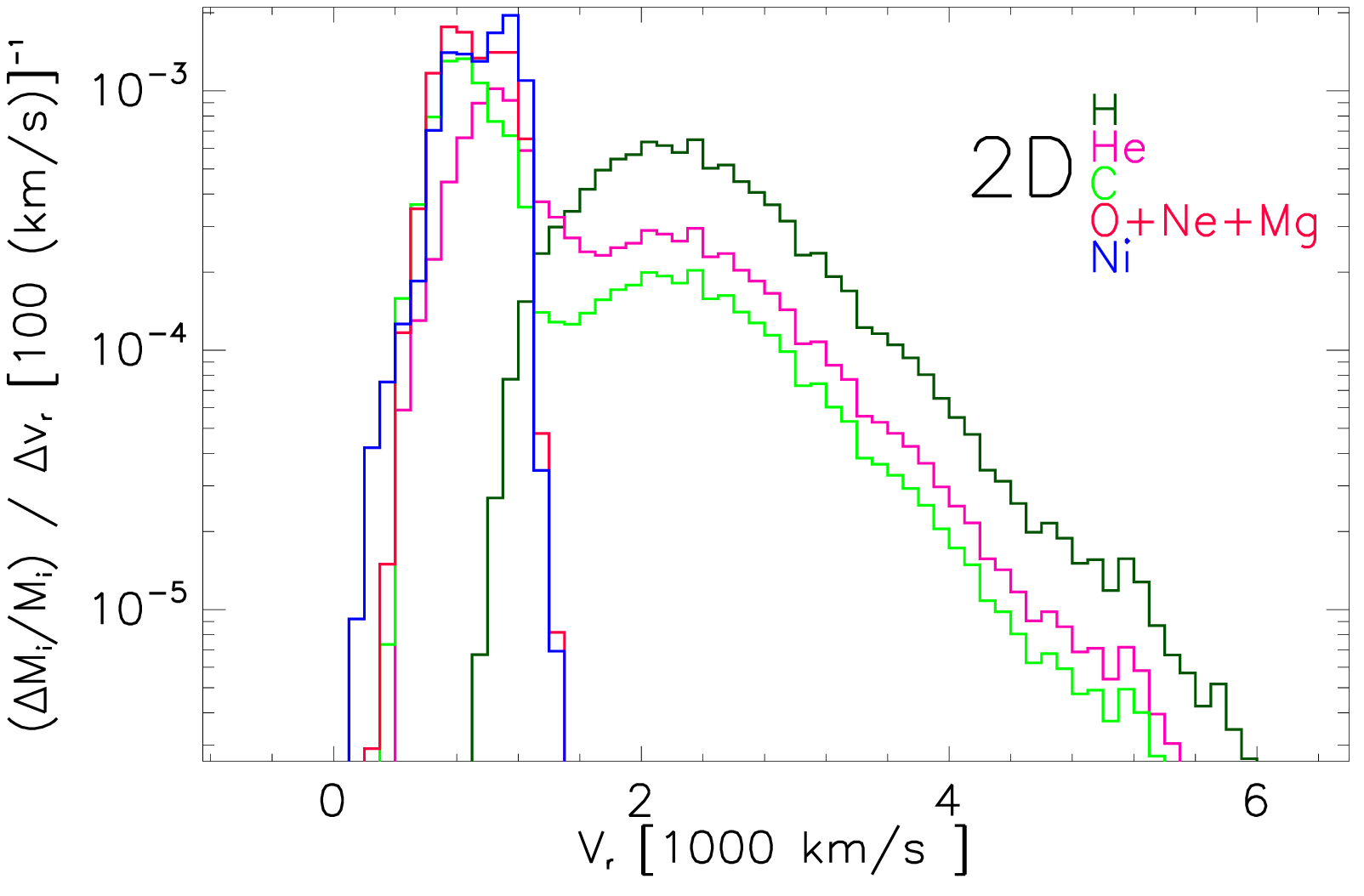}
\includegraphics[angle=90,width=.40\textwidth]{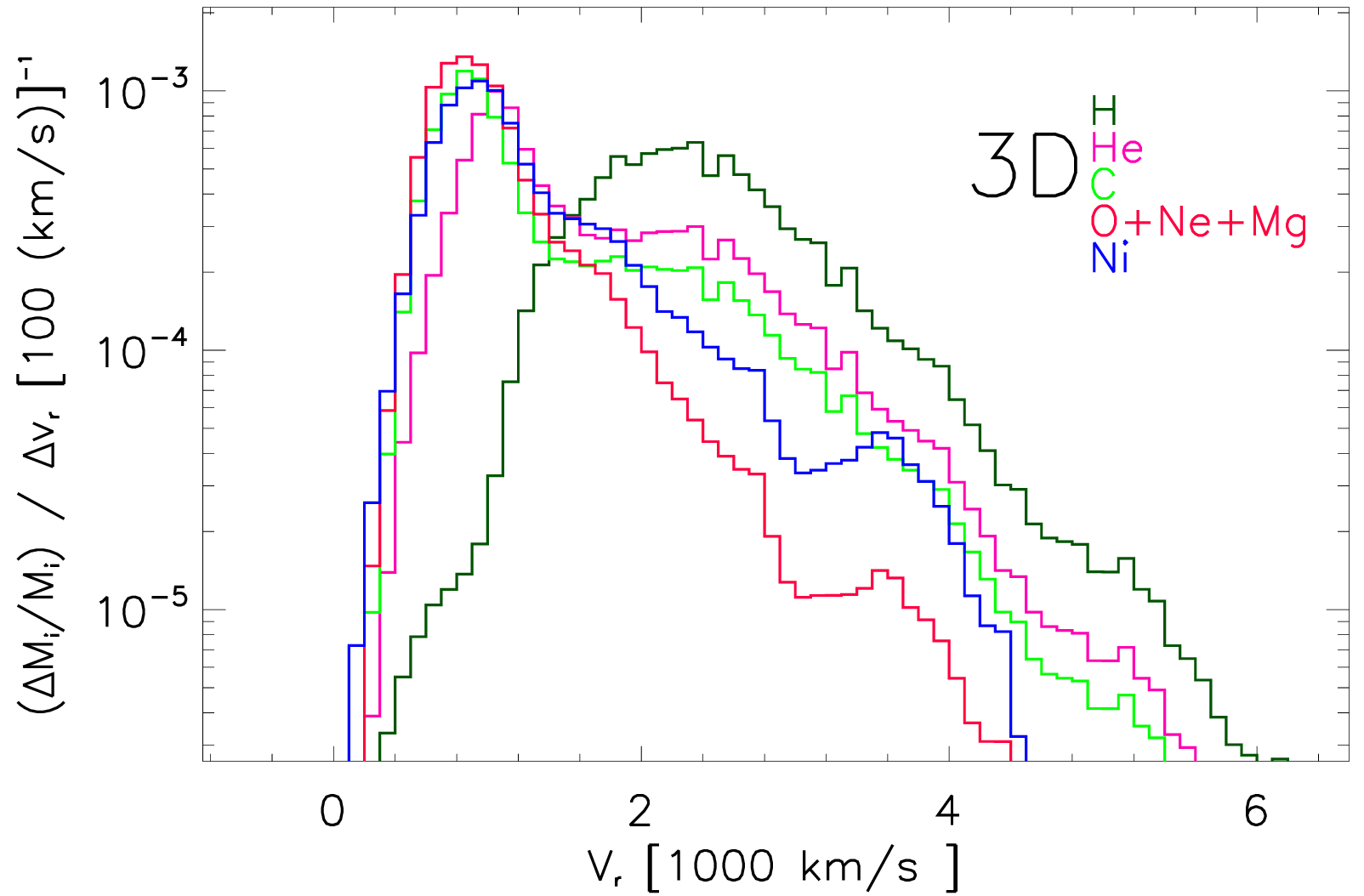}
\caption{\label{fig:3D2Dvsvel}
  Normalized mass distributions of hydrogen (black), helium (magenta),
  carbon (green), oxygen plus neon and magnesium (red), 
  and ``nickel'' (including iron-group elements and silicon; blue)
  versus radial velocity $v_r$ for the 3D simulation (right) and a
  2D simulation performed for the meridional slice of the 3D model
  indicated in Fig.~\ref{fig:2Dcut} (left).
  From top to bottom, the distributions
  are given at about 350$\,$s, 1000$\,$s, 2600$\,$s, and 9000$\,$s
  after core bounce. The binning is done in intervals of $\Delta v_r =
  100\,$km$\,$s$^{-1}$ and the distributions $\Delta M_i/M_i$ with $i$
  being the element index are given per unit length of velocity. Note
  the large differences between the 3D and 2D results of the O and Ni
  distributions at high velocities and of the hydrogen distribution at
  low velocities.}
\end{center}
\end{figure*}
%%%%%%%%%%%%%%%%%%%%%%%%%%%%%%%%%%%%%%%%%%%%%%%%%%%%%%%%

%%%%%%%%%%%%%%%%%%%%%%%%%%%%%%%%%%%%%%%%%%%%%%%%%%%%%%%%
%% FIGURE 7
%%%%%%%%%%%%%%%%%%%%%%%%%%%%%%%%%%%%%%%%%%%%%%%%%%%%%%%%
\begin{figure*}
  \begin{center}
\includegraphics[angle=90,width=.40\textwidth]{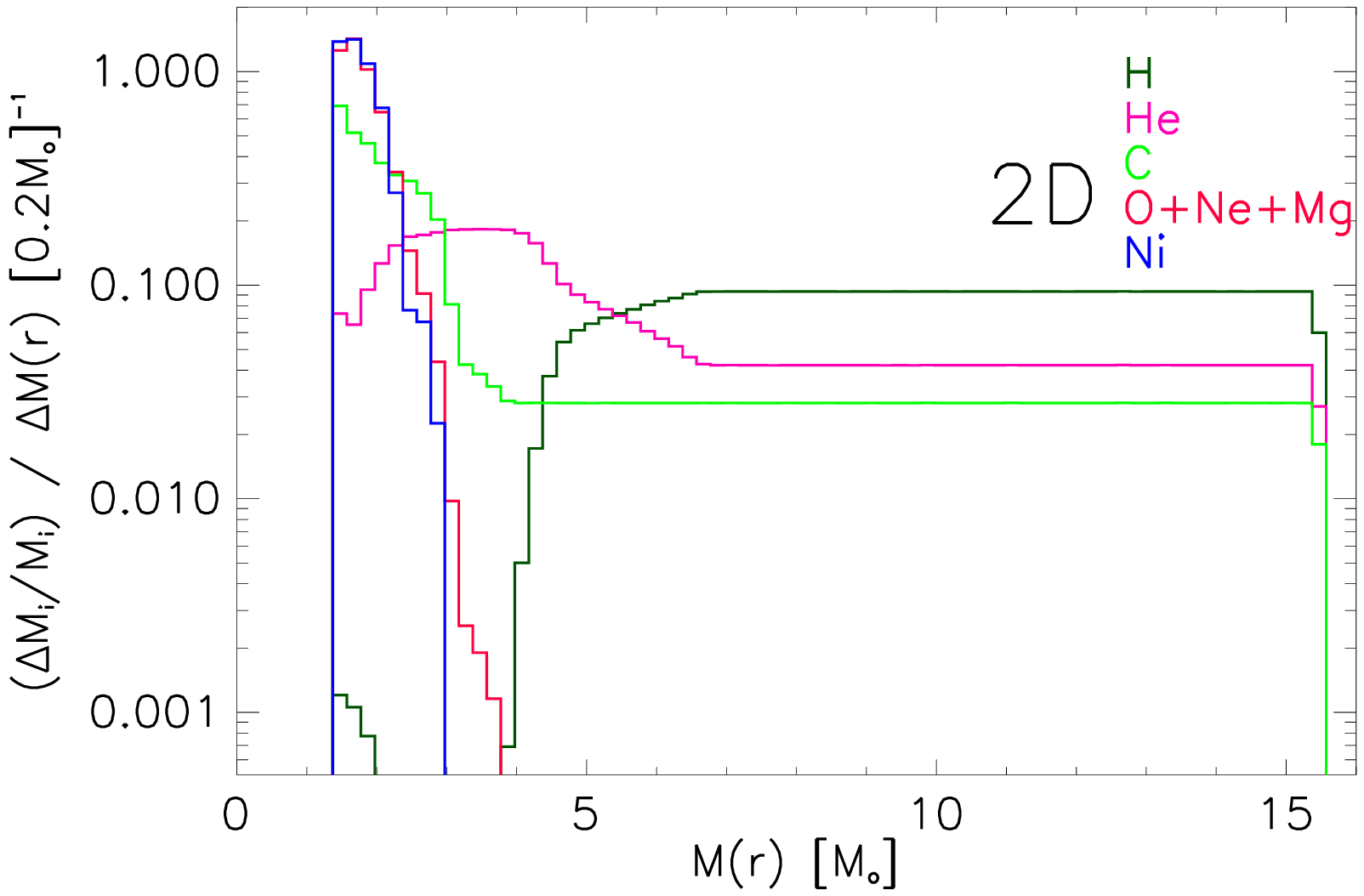}
\includegraphics[angle=90,width=.40\textwidth]{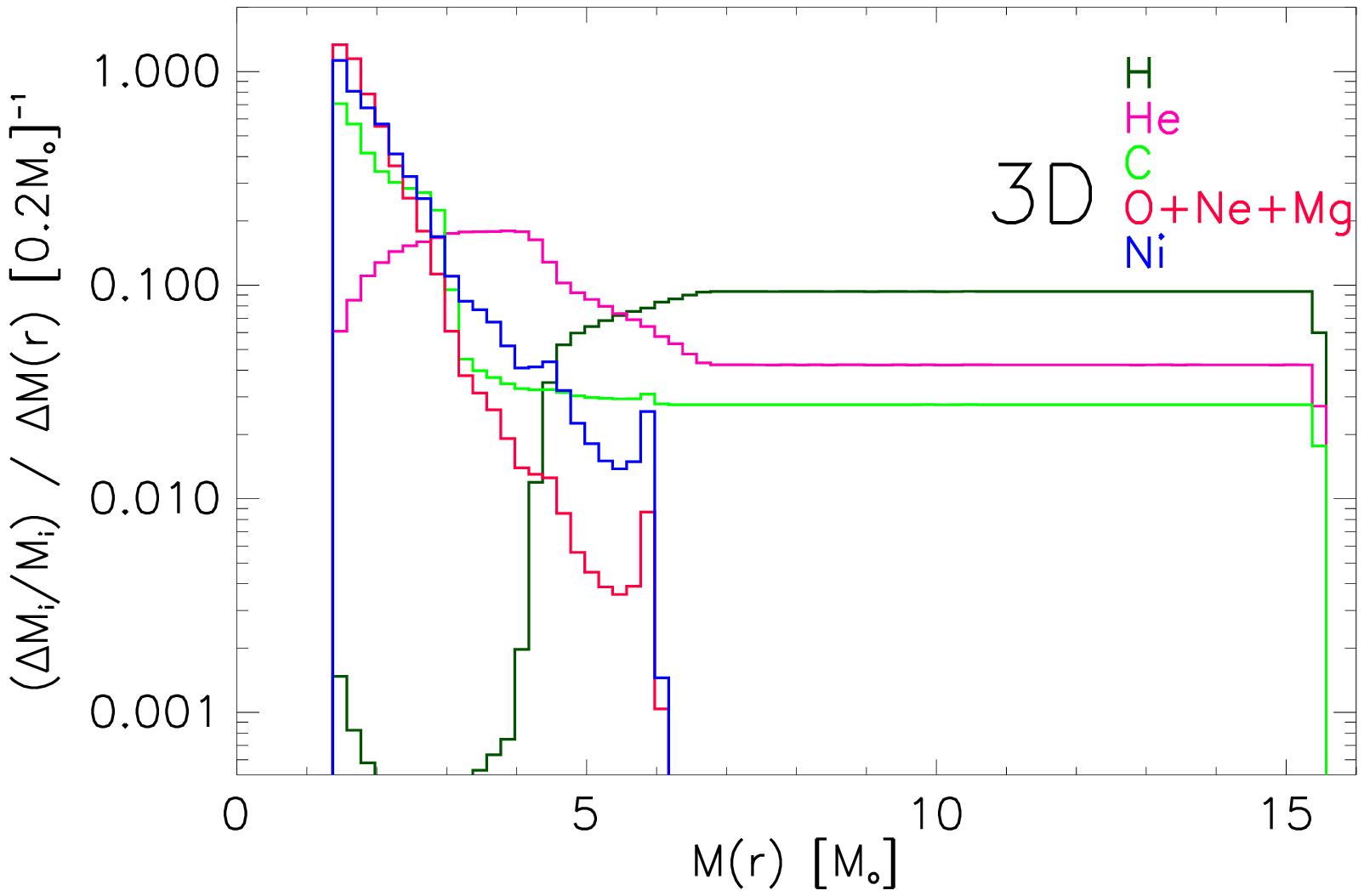}\\
\includegraphics[angle=90,width=.40\textwidth]{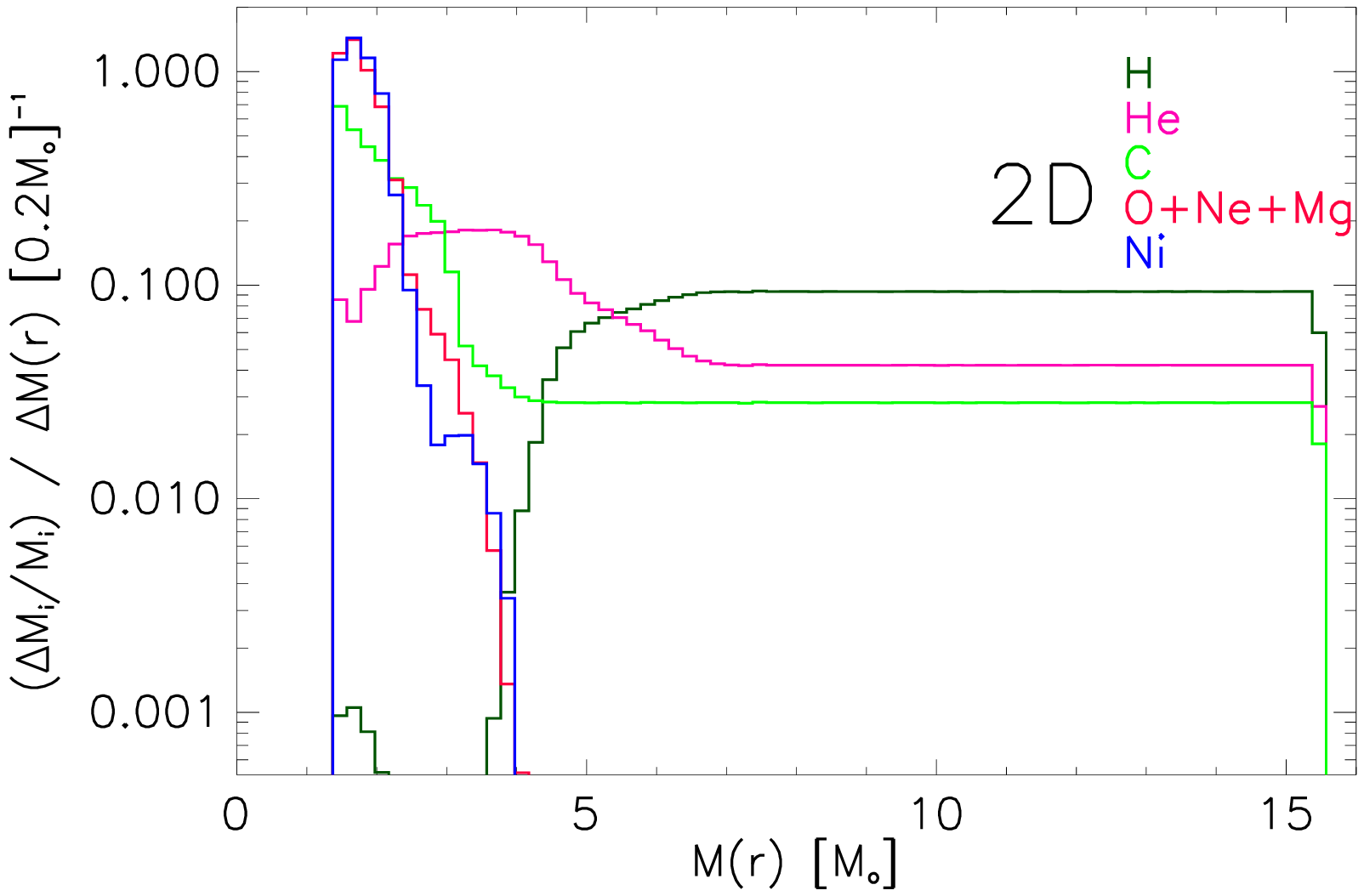}
\includegraphics[angle=90,width=.40\textwidth]{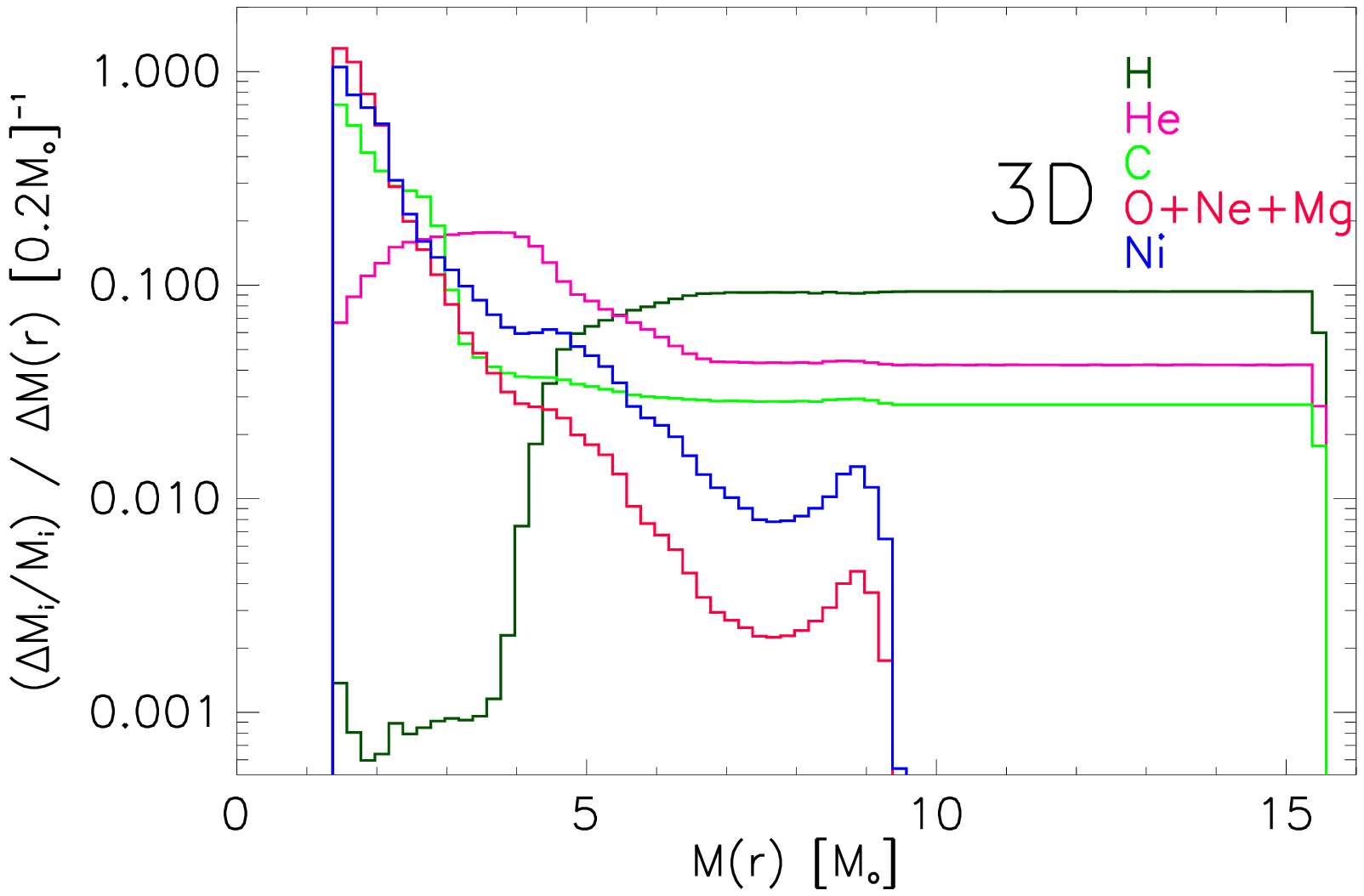}\\
\includegraphics[angle=90,width=.40\textwidth]{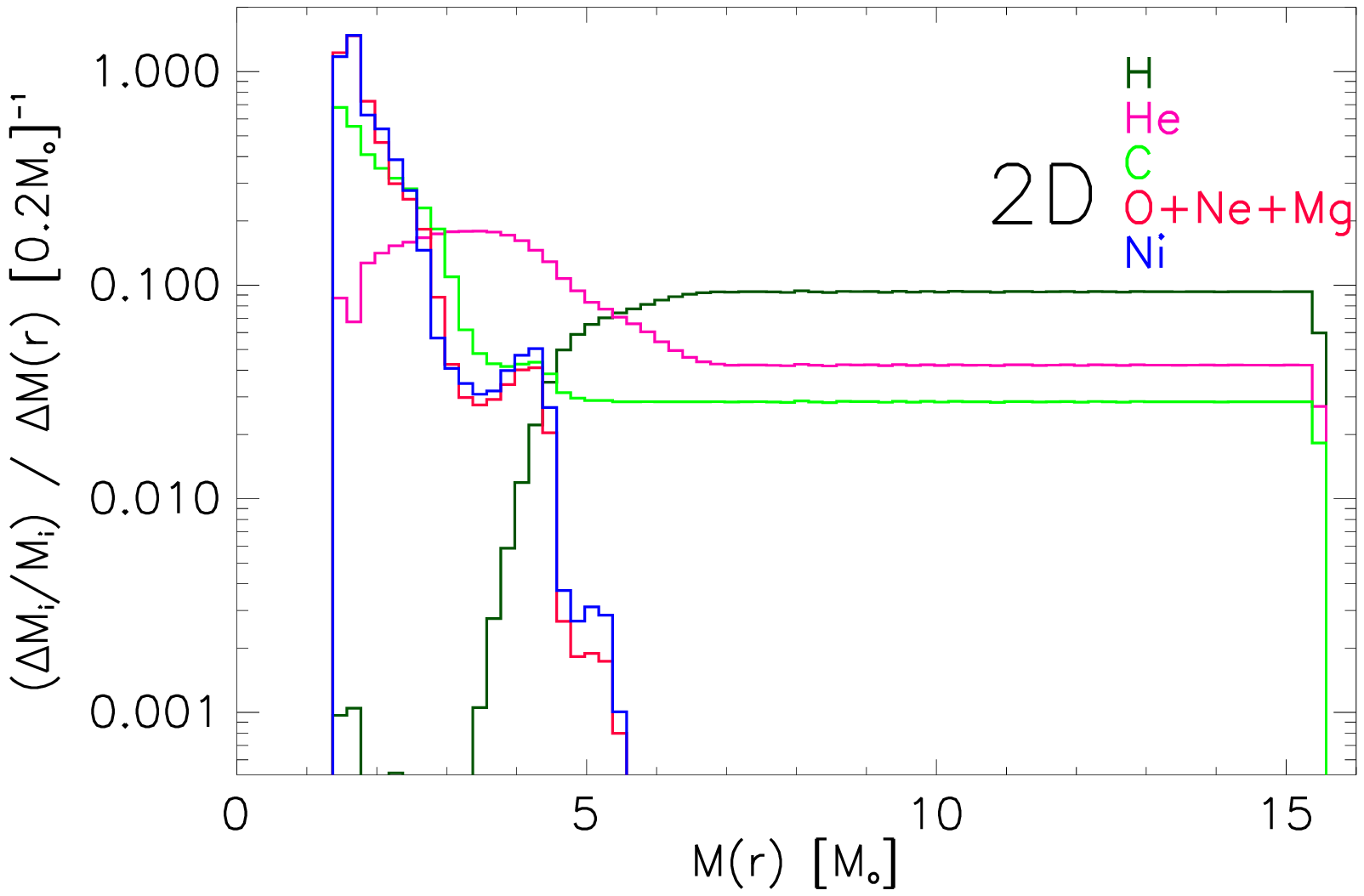}
\includegraphics[angle=90,width=.40\textwidth]{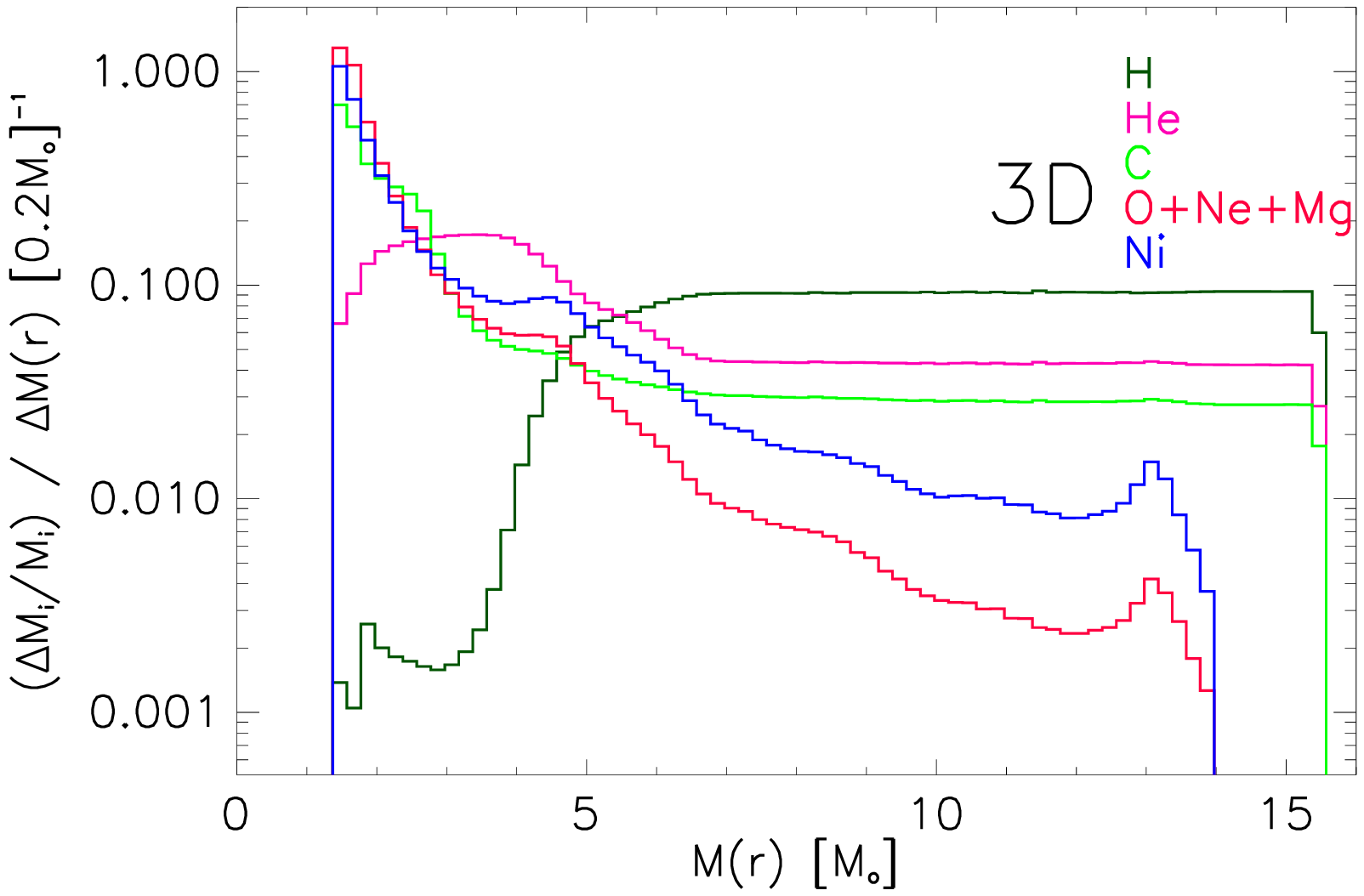}\\
\includegraphics[angle=90,width=.40\textwidth]{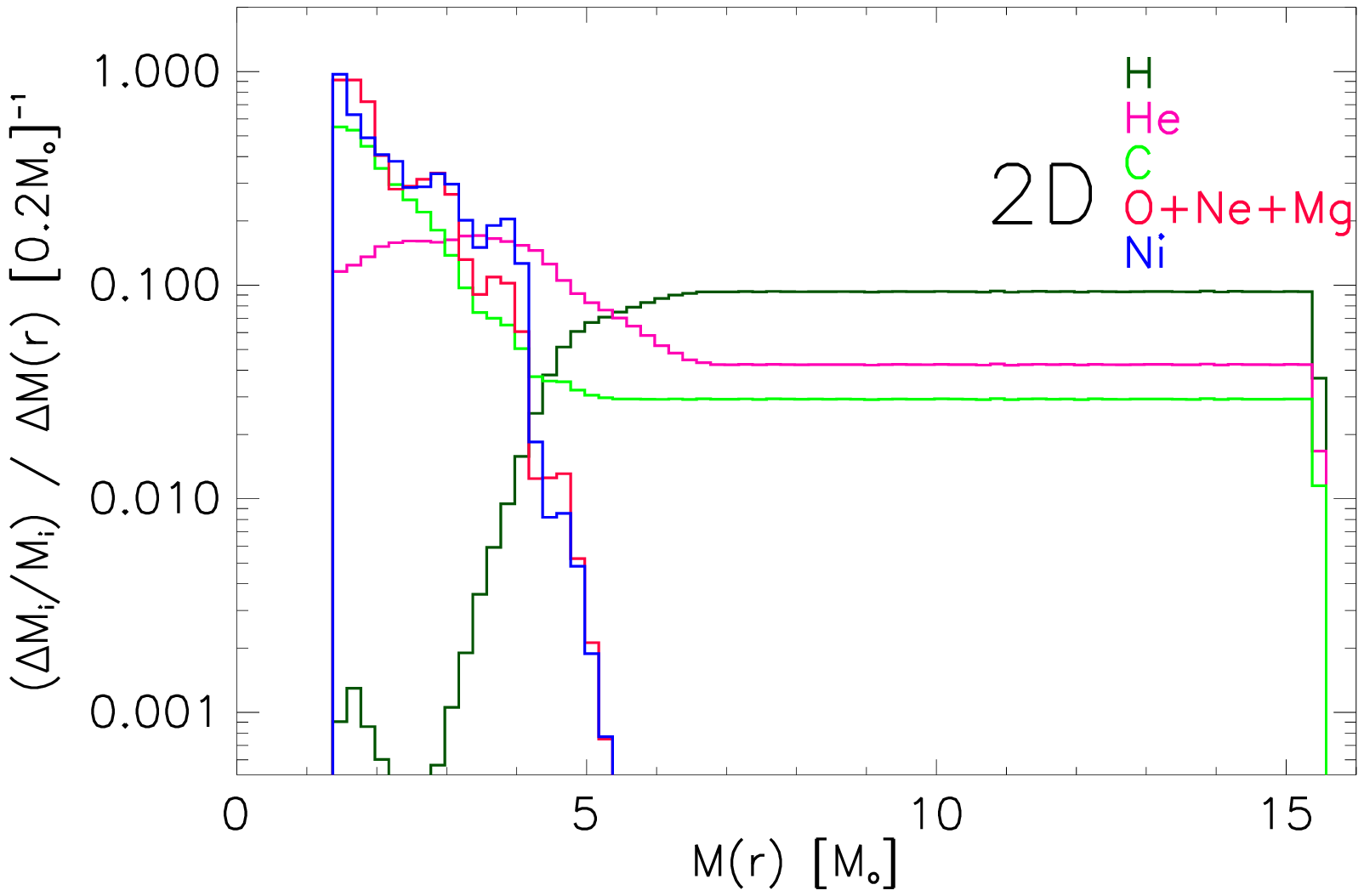}
\includegraphics[angle=90,width=.40\textwidth]{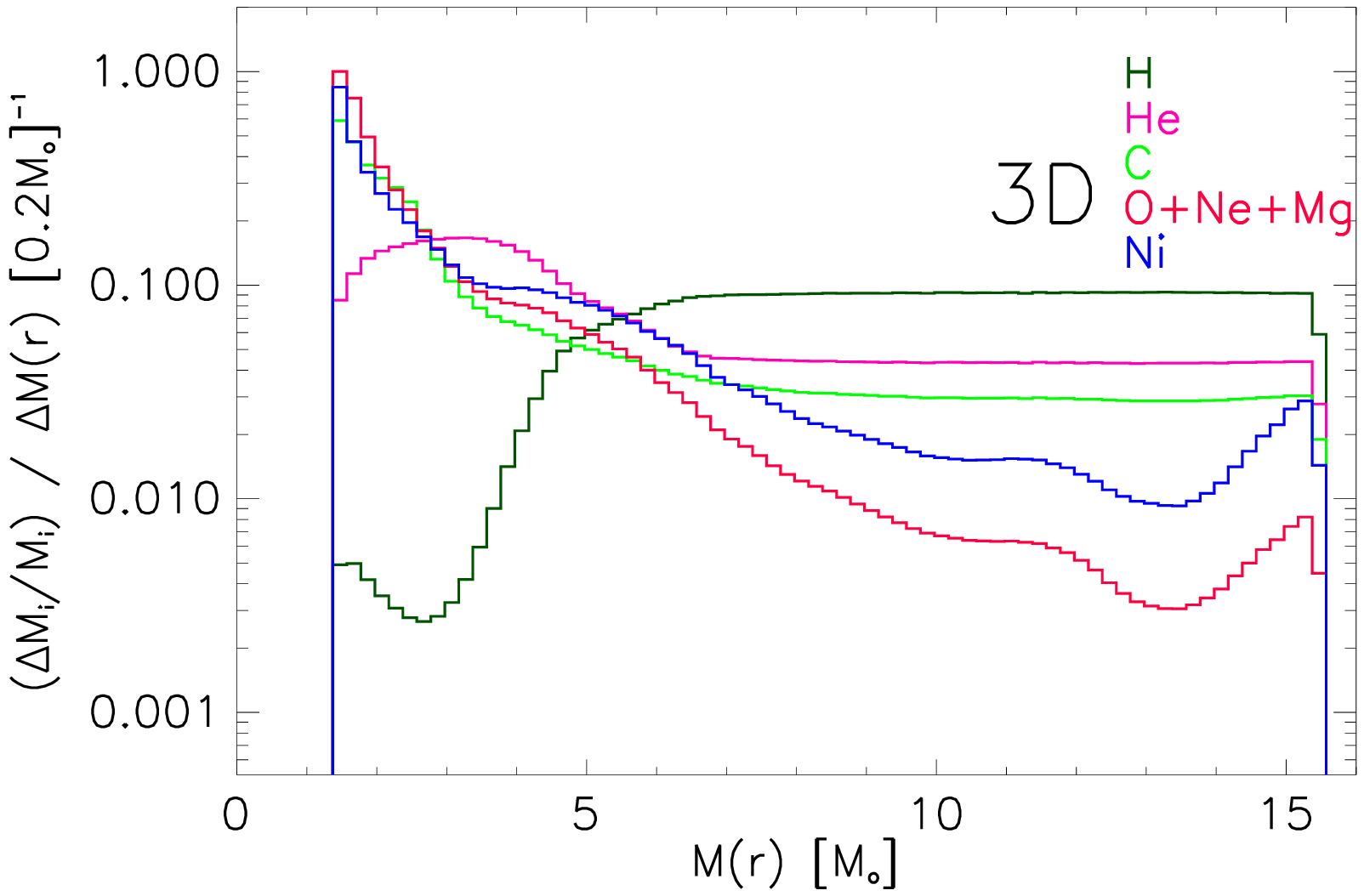}
\caption{\label{fig:3D2Dvsmass}
  Normalized mass distributions of hydrogen (black), helium (magenta),
  carbon (green), oxygen plus neon and magnesium (red),
  and ``nickel'' (including iron-group elements and silicon; blue)
  versus enclosed mass $M(r)$ for the 3D simulation (right) and a
  2D simulation performed for the meridional slice of the 3D model
  indicated in Fig.~\ref{fig:2Dcut} (left).
  From top to bottom, the distributions
  are given at about 350$\,$s, 1000$\,$s, 2600$\,$s, and 9000$\,$s
  after core bounce. The binning is done in intervals of $\Delta M(r) =
  0.2\,M_\odot$ and the distributions $\Delta M_i/M_i$ with $i$ being
  the element index are given per unit length of mass. Note that in 3D
  oxygen and nickel clumps are able to penetrate through the hydrogen
  envelope, while they are stuck in the helium layer in the 2D case.}
\end{center}
\end{figure*}
%%%%%%%%%%%%%%%%%%%%%%%%%%%%%%%%%%%%%%%%%%%%%%%%%%%%%%%%

\subsection{Radial element mixing}
\begin{figure}
\begin{center}
\includegraphics[angle=90,width=.48\textwidth]{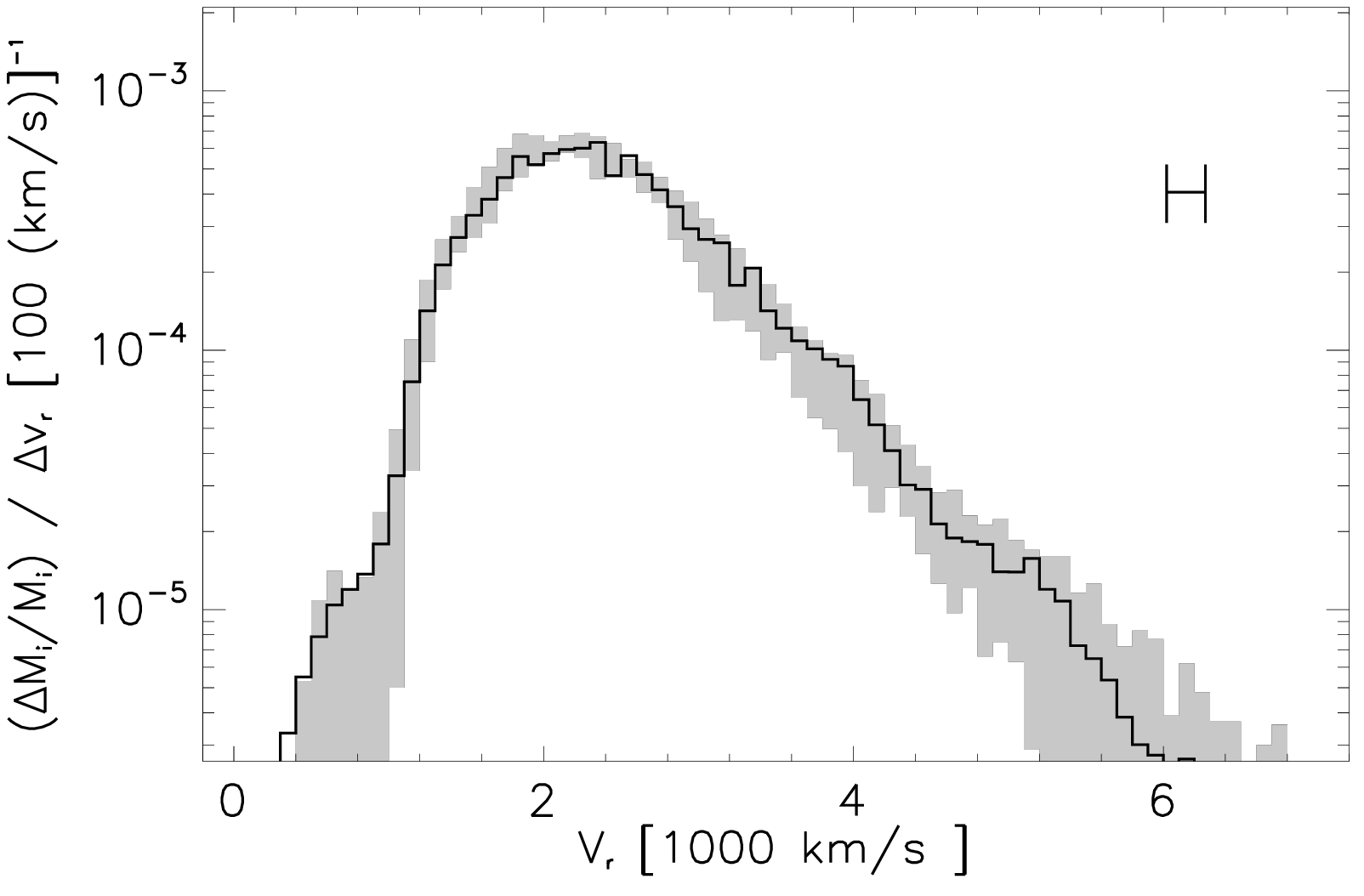}\\[2mm]
\includegraphics[angle=90,width=.48\textwidth]{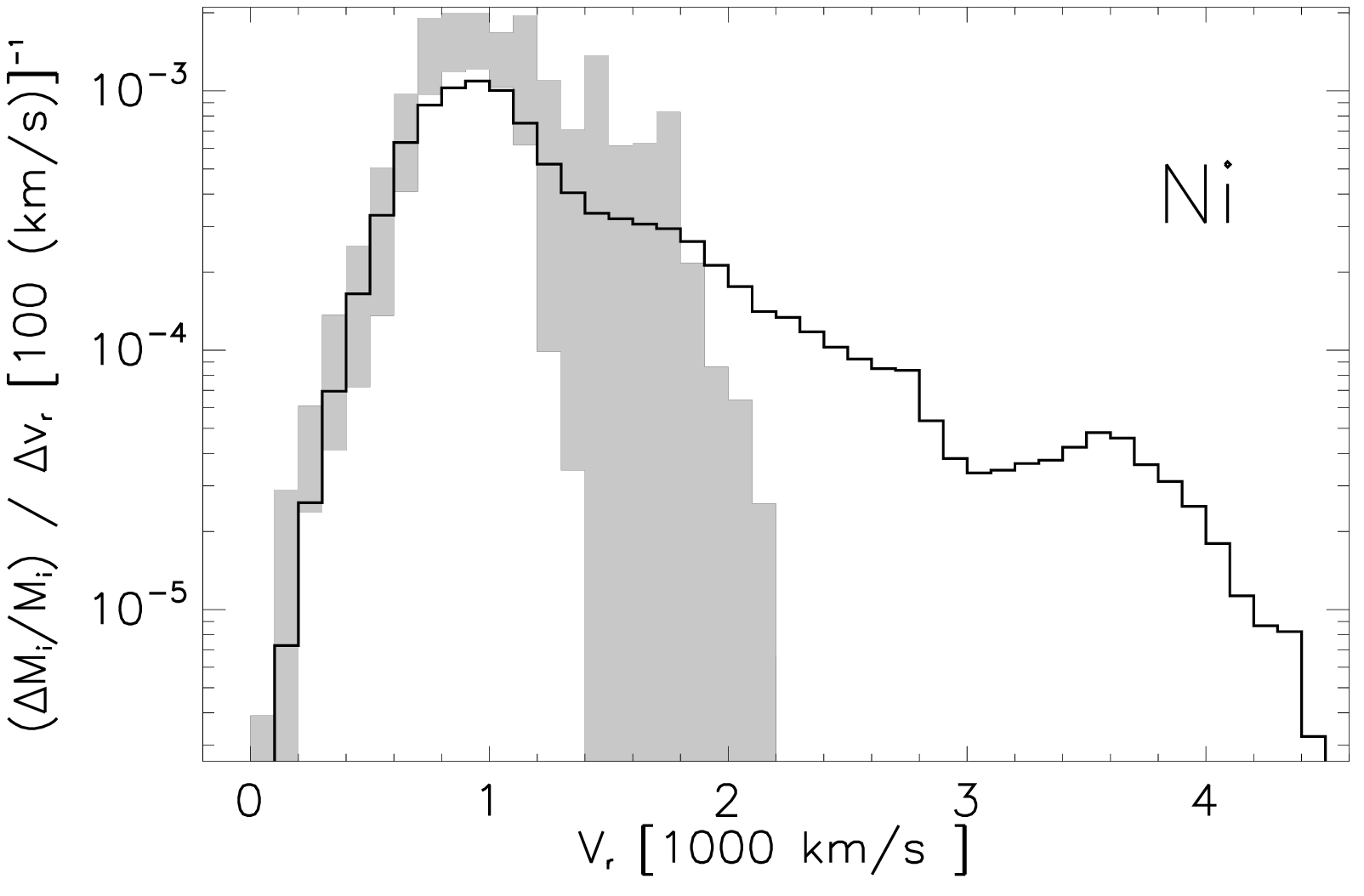}\\[2mm]
\includegraphics[angle=90,width=.48\textwidth]{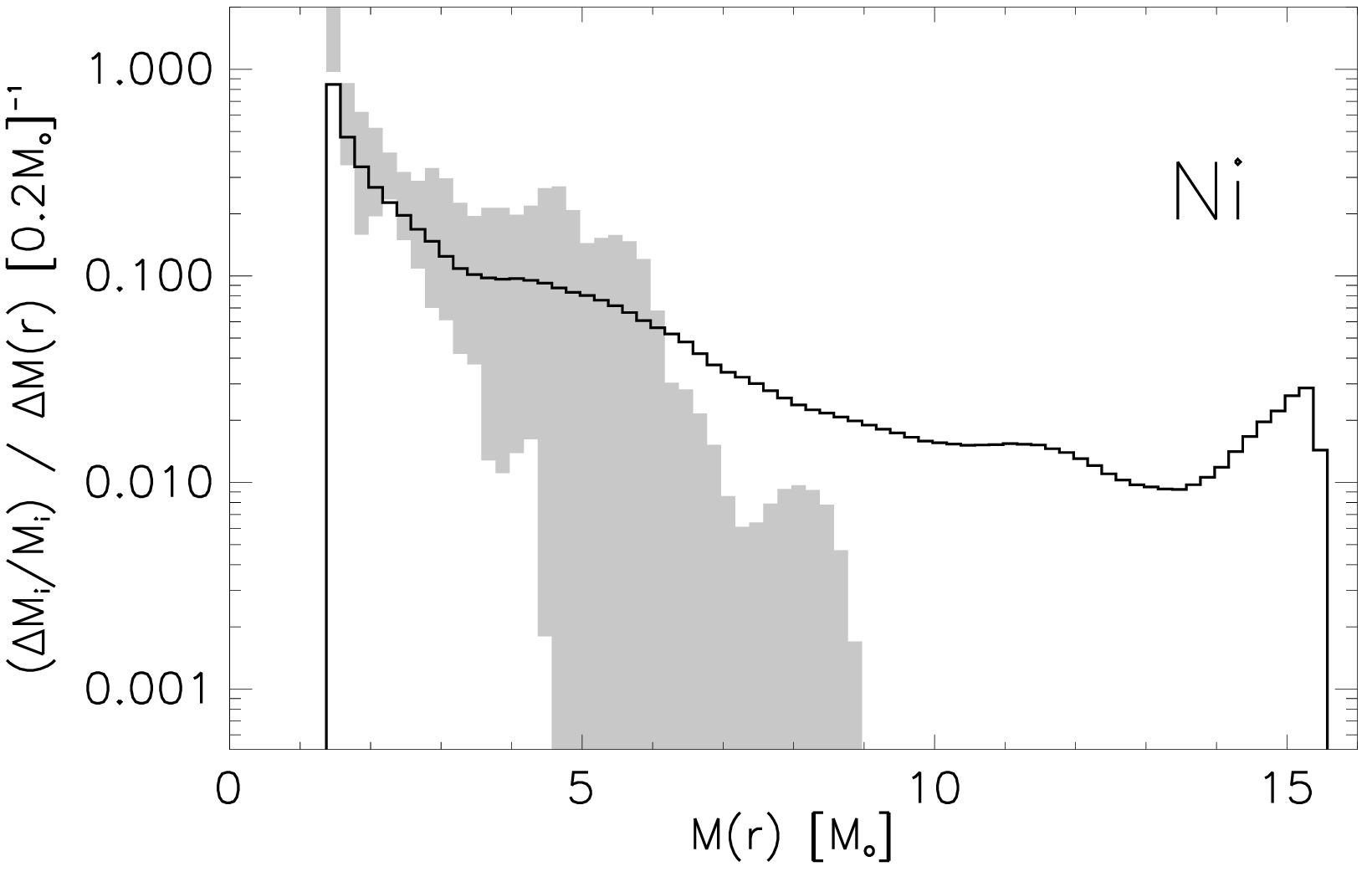}
\caption{\label{fig:HNivsvel}
  Normalized mass distributions of hydrogen ({\em upper panel}) 
  versus radial velocity $v_r$ and of
  ``nickel'' (including iron-group elements and silicon)
  versus radial velocity $v_r$ ({\em middle panel}) as well as
  versus enclosed mass $M(r)$ ({\em bottom panel}).
  Results are shown for the 3D simulation (solid line)
  compared to the corresponding 2D results (grey region) at about
  9000$\,$s. The binning is done in intervals of $\Delta v_r =
  100\,$km$\,$s$^{-1}$ and of $\Delta M(r) = 0.2\,M_\odot$, 
  respectively, and the distributions $\Delta M_i/M_i$ with $i$
  being the element index are given per unit length of velocity
  or mass, respectively. The
  grey-shaded area indicates the significant variation between the 
  thirteen 2D simulations, which were started from different meridional
  slices of the initial 3D explosion model.}
\end{center}
\end{figure}

Fig.~\ref{fig:3Dsurfaces} displays the development of these
fast-moving clumps during our 3D explosion run by showing 
surfaces of constant mass fractions 
of carbon, oxygen, and nickel for two different viewing
directions and two different times (350$\,$s and $\sim$9000$\,$s after
core bounce). Figure~\ref{fig:3Dvolume} provides a volume-rendered
image of the composition distribution at the later time, while
Fig.~\ref{fig:3Dcuts} gives composition information on cut planes 
through the mixed stellar core and some of the major plumes of 
different types. Finally, 
Figs.~\ref{fig:3D2Dvsvel} and \ref{fig:3D2Dvsmass} present normalized
mass distributions of various nuclear species in the radial-velocity
and enclosed-mass space for our 3D simulation compared to an
``average'' 2D result at several times after core bounce, and 
Fig.~\ref{fig:HNivsvel} provides information for the spread of the
hydrogen and nickel mass distributions in our set of 2D runs at the
end of the simulations.

We stress that what we denote as ``nickel'' here and in the 
following actually includes
the contributions of silicon and of the iron-group elements from 
the postshock region. This makes sense
because the shock front in the initial model has nearly reached the
oxygen layer (which implies that only a very small fraction of the 
original silicon is left) and because we do not account for 
nuclear burning in our simulations (for which reason no silicon is
produced lateron); see Sect.~\ref{sec:eos}. In the
3D model the total mass of free neutrons is $6\times
10^{-4}\,M_\odot$, of hydrogen 8.16$\,M_\odot$, of helium
5.40$\,M_\odot$, of carbon 0.12$\,M_\odot$, of oxygen 0.20$\,M_\odot$,
of neon 0.048$\,M_\odot$, of magnesium 0.005$\,M_\odot$, and of
``nickel'' 0.212$\,M_\odot$. About 1.35$\,M_\odot$ of the
15.5$\,M_\odot$ progenitor are thus absorbed into the neutron star. In
the 2D runs based on selected meridional slices, the ejecta masses of
the different nuclear species can differ insignificantly from the
listed numbers.

Figure~\ref{fig:3Dsurfaces} shows that at 350$\,$s oxygen and nickel
rich, dense fingers, some of which display the typical mushroom shape
of Rayleigh-Taylor structures, penetrate through the outer boundary of
the carbon layer of the exploding star. The corresponding $^{12}$C
surface is defined by a mass fraction of 3\% (note that also exterior
to this surface, in the surrounding helium and hydrogen layers, a
small but non-zero mass fraction of carbon is still present and 
representative of the heavy elements that define the metallicity of
the progenitor star). The most prominent features at that time 
have been seeded by the biggest high-entropy bubbles visible in
Fig.~\ref{fig:inimod} and develop to the largest mushrooms visible
more than 8000$\,$s later. 

The chemical composition of the various kinds of extended 
plumes in this late stage is very interesting. All of them
contain high mass fractions of hydrogen and helium,
which account for about 50 to 70 percent of their mass with spatial
variations in individual filaments. The dominant heavy species, 
however, differs between different structures. The largest mushrooms
carry predominantly nickel and may contain
a narrow spine with oxygen. In contrast, many of the far more
numerous, less extended and more narrow fingers are composed of a
mixture of nickel and oxygen. Also clumps with a dominant mass
fraction of oxygen can be identified. More useful for such an analysis
than the surface images of Fig.~\ref{fig:3Dsurfaces} is the
visualization by volume-rendering techniques in Fig.~\ref{fig:3Dvolume}.
While the surfaces of constant mass fraction can be misleading,
because their shape depends strongly on the chosen value and no
information about the interior composition is provided, the 
ray-tracing image shows that the largest Rayleigh-Taylor mushrooms
carry mostly nickel, the majority of the more extended fingers
are composed of a mix of elements, and only some clumpy, relatively
compact, knotty features (clearly visible along the edge of the 
pastel-colored and whitish glowing core in Fig.~\ref{fig:3Dvolume})
contain a dominant fraction of oxygen. This is confirmed by the 
planar cuts through two typical Rayleigh-Taylor structures in
Fig.~\ref{fig:3Dcuts}. The upper left panel of this figure, in 
which a 2D cut through the whole core is displayed, demonstrates
that on the global scale there are extended and separated regions 
where different species are the most abundant chemical constituent.
In the core the hydrogen and helium admixture is much smaller than
in the extended fingers and plumes. Except in some smaller ``bubbles'',
where H+He dominate, heavies contribute 50--80\%, in some clumps 
even more than 90\%, of the core composition.

The time sequences of the normalized mass distributions of the 
nuclear species in the radial-velocity
and enclosed-mass space\footnote{The mass distribution of each nuclear
  species is normalized such that the radial integration in velocity
  or enclosed-mass space gives unity. Thereby the mass $\Delta M_i$
  of an element $i$ in a velocity or mass bin is divided by the
  total mass $M_i$ of this element on the grid at the time of
  evaluation.} of Figs.~\ref{fig:3D2Dvsvel} and \ref{fig:3D2Dvsmass}
visualize the gradual acceleration of the outer
He and H layers of the exploding star by the outgoing shock front.
Simultaneously, the initially fast metals of the core penetrate into
the outer composition layers and are decelerated.
Both the deceleration and the mixing depend
crucially on the hydrodynamic instabilities that develop at the
composition interfaces, seeded by the initial explosion asymmetries.
At 350$\,$s the helium ``wall'' can be discerned as the peak of the 
distribution between 3500$\,$km$\,$s$^{-1}$ and 4500$\,$km$\,$s$^{-1}$
at an enclosed mass between $\sim$2.5$\,M_\odot$ and 
$\sim$4.5$\,M_\odot$.
From a comparison of the plots for 2D and 3D it is obvious that 
nickel and oxygen experience a much stronger deceleration during
the subsequent evolution and far
less efficient mixing into the hydrogen envelope in the 2D
case (the plane of the meridional cut considered in 
Figs.~\ref{fig:3D2Dvsvel} and \ref{fig:3D2Dvsmass} is
displayed in Fig.~\ref{fig:2Dcut})\footnote{The set of 
  meridional slices of the 3D initial data
  that we used for 2D simulations exhibits variations at a certain
  level with respect to gas energies and asymmetries of the mass
  distribution. This leads to corresponding differences in the
  development of nonradial hydrodynamic instabilities in the 
  outer stellar layers. The case picked for
  Figs.~\ref{fig:3D2Dvsvel} and \ref{fig:3D2Dvsmass} can be
  considered as ``average'' concerning metal clump velocities and
  mixing, but some of the 2D slices also yield slightly more extreme
  results (cf.\ Fig.~\ref{fig:HNivsvel}). In particular, the
  set of slicing directions was chosen such that 3D regions with 
  particularly strong growth of Rayleigh-Taylor fingers were not 
  missed (Sect.~\ref{sec:simulations}).  }.
Since neon and magnesium are carbon-burning ashes and
well mixed with oxygen in the oxygen-layer of the progenitor
star, the distributions of these three elements closely match
each other during the mixing evolution and can be hardly 
distinguished on plots. We therefore have combined them into
one line in Figs.~\ref{fig:3D2Dvsvel} and \ref{fig:3D2Dvsmass}.

%%%%%%%%%%%%%%%%%%%%%%%%%%%%%%%%%%%%%%%%%%%%%%%%%%%%%%%%
%% FIGURE 8
%%%%%%%%%%%%%%%%%%%%%%%%%%%%%%%%%%%%%%%%%%%%%%%%%%%%%%%%

%%%%%%%%%%%%%%%%%%%%%%%%%%%%%%%%%%%%%%%%%%%%%%%%%%%%%%%%

A closer inspection of Figs.~\ref{fig:3D2Dvsvel} and
\ref{fig:3D2Dvsmass} reveals that the metal distributions in
both 2D and 3D models peak at the location of the metal core 
(enclosed mass $M_r \la 2.5\,M_\odot$) with similar velocities
(around 2000--3000$\,$km$\,$s$^{-1}$
at 350$\,$s and 1000$\,$s, around 1500--2500$\,$km$\,$s$^{-1}$
at 2600$\,$s, and around 1000$\,$km$\,$s$^{-1}$ at 9000$\,$s). 
In 3D, however, the nickel and oxygen distributions possess
much more massive and wider high-velocity tails up to 
velocities of more than 6000$\,$km$\,$s$^{-1}$ at 350$\,$s.
In the long run this is causal for
the large discrepancies of the radial composition mixing.
Two effects appear to contribute to the formation and continued
presence of these high-velocity tails:
\begin{itemize}
\item[(i)]
In 3D the Rayleigh-Taylor instabilities at the composition interfaces
(in particular the (C+O)/He boundary) grow more rapidly. This
leads to the formation of extremely fast (mostly nickel and less
oxygen) clumps with velocities in excess of 5000$\,$km$\,$s$^{-1}$
at 350$\,$s (upper right panel of Fig.~\ref{fig:3D2Dvsvel}). These
dense bullets penetrate deep into the hydrogen layer even before
the helium wall has formed (see the upper two right panels in 
Fig.~\ref{fig:3D2Dvsmass}). The more rapid growth of the 
Rayleigh-Taylor instabilities also seems to be the reason for
the efficient mixing of hydrogen into the metal core, which is
clearly different from the 2D case already at $\sim$1000$\,$s.
\item[(ii)]
A larger fraction of the nickel and oxygen with a velocity of
$\sim$4000$\,$km$\,$s$^{-1}$ at 350$\,$s, which is in the ballpark of
the maximum nickel and oxygen velocities of the 2D models, experiences 
less velocity reduction in the 3D simulation. This is obviously 
linked to a different propagation and deceleration of the dense
metal-containing clumps as they move quasi-ballistically through the
helium shell and inner part of the hydrogen envelope under the 
influence of drag forces exerted by the
surrounding medium. While the fastest bullets with a speed of
5000--6000$\,$km$\,$s$^{-1}$ at 350$\,$s still move with
3000--4500$\,$km$\,$s$^{-1}$ at $t > 2600\,$s, the clumps with 
initially $\sim$4000$\,$km$\,$s$^{-1}$ are decelerated to 
2000--3000$\,$km$\,$s$^{-1}$. In contrast, in 2D none of the 
material of the metal core has retained velocities of more than
2000$\,$km$\,$s$^{-1}$ at the time when the shock reaches the
stellar surface ($t \sim 8000\,$s). 
We will return to a detailed discussion 
of this very interesting phenomenon in Sect.~\ref{sec:clumpprop},
where we will present simple analytic arguments for an explanation.
\end{itemize}
 
In Fig.~\ref{fig:HNivsvel} the normalized mass distributions of
hydrogen and nickel in the radial velocity space are plotted at 
$t \sim 9000\,$s shortly after the shock breakout from the stellar
surface (when the expansion is close to become homologous) 
for our 3D run in comparison to the compilation of
results obtained for all 2D simulations with different chosen
meridional slices. While the maximum hydrogen
velocities are very similar in 2D and 3D and reach up to about
6000$\,$km$\,$s$^{-1}$, a clearly bigger hydrogen mass attains the 
lowest velocities (below 1000$\,$km$\,$s$^{-1}$) in the 3D case.
Correspondingly, a small fraction of the hydrogen (some hundredths
of a solar mass) is mixed deep into the metal core of the star
to regions with enclosed masses of $M(r)\la 2\,M_\odot$
(see also the right panels of Fig.~\ref{fig:3D2Dvsmass}).  
These findings are very similar to the results
obtained in 2D by Kifonidis et al. (2006; cf.\ figures 6, 8, and 9
there), who considered cases with largely aspherical beginning of the
explosion as discussed in Sect.~\ref{sec:dynevol}). 
In 3D similarly strong inward
mixing of hydrogen does not require a very pronounced global asymmetry
of the early blast.  Richtmyer-Meshkov instability at the
shock-distorted He/H composition interface
plays no crucial role in our
3D simulations, because the outgoing supernova shock is significantly
less deformed than in the 2D models discussed by
\citet{kifonidis:2006a}, see Sect.~\ref{sec:dynevol}.

Even more impressive than the hydrogen differences is the large
discrepancy of 2D and 3D with respect to the nickel distribution in
velocity space and the maximum nickel velocities.  While in 2D even
the fastest nickel clumps move with just 2200$\,$km$\,$s$^{-1}$
at $t \sim 9000\,$s (Fig.~\ref{fig:HNivsvel}), a large fraction 
of the nickel retains velocities
of more than 3000$\,$km$\,$s$^{-1}$ and some even more than
4000$\,$km$\,$s$^{-1}$ in 3D (Fig.~\ref{fig:HNivsvel}).  From
Fig.~\ref{fig:3D2Dvsvel} we can conclude that similarly large 2D
vs.\ 3D differences are obtained for the oxygen distribution in
velocity space, whereas the hydrogen and helium distributions of the
2D case are close to the results of the 3D run (except for the
low-velocity tail of the hydrogen distribution mentioned above), and
the carbon distribution exhibits slight differences only between about
1000$\,$km$\,$s$^{-1}$ and 2500$\,$km$\,$s$^{-1}$. Interestingly, a
significantly larger fraction of the nickel than of the oxygen 
propagates with the highest speeds of clumps. This is compatible with
the fact that the metal content of the biggest and most extended 
Rayleigh-Taylor structures in Fig.~\ref{fig:3Dsurfaces} is
dominated by nickel, while oxygen is concentrated mostly in  
thin, shorter cores of the mushrooms (see Fig.~\ref{fig:3Dcuts}).

It should also be noted that the spread of the maximum nickel
velocities obtained in the set of 2D runs in Fig.~\ref{fig:HNivsvel} 
is considerably smaller than the corresponding spread of the 
2D results in the mass space. The maximum Ni velocities vary
only between about 1700$\,$km$\,$s$^{-1}$ for average and
2200$\,$km$\,$s$^{-1}$ for extreme 2D cases (in which the meridional
slices were placed at the locations of the biggest nickel mushrooms
in 3D). In the former models nickel is transported to slightly
more than 5$\,M_\odot$, whereas in the latter ones some of the
nickel gets mixed out to an enclosed mass of even 8--9$\,M_\odot$.
In all the cases, however, the mixing of iron-group elements into
the hydrogen layer is much less strong in 2D than in 3D.

Figure~\ref{fig:3D2Dvsmass}, showing an average 2D run, confirms 
these findings. In the latest stages nickel and oxygen can be 
observed at an enclosed mass of about
5$\,M_\odot$ (in rough agreement with the results of
\citet{kifonidis:2003a}, see their fig.~15), which is close to the
base of the hydrogen envelope\footnote{The slightly 
deeper intrusion of metals into hydrogen in our models can be 
understood as a consequence of the lower explosion energy, which 
leads to a slower shock propagation and later formation of the 
helium wall than in the simulations of \citet{kifonidis:2003a}.}.
In typical 2D cases the metal-containing clumps thus
get stuck in the dense helium wall as pointed out by
\citet{kifonidis:2003a}. In contrast, in the 3D run nickel and oxygen
are mixed deep into the hydrogen layer (and also a small amount of
carbon is carried from the metal core into the hydrogen shell). 
The fastest nickel and
oxygen clumps are able to penetrate nearly to the surface of the
exploding star (in the mass space).

The two humps that can be discriminated in the mass distribution of
these nuclear species well ahead of the bulk of the metal core in the
mass and velocity space at 9000$\,$s, are linked to the few, but big
and far-reaching structures, and to the greater number of smaller
fingers that extend more uniformely to all directions in
Fig.~\ref{fig:3Dsurfaces}. The bigger bullets contain about three
to four times more nickel than oxygen. They  
carry in total a mass of $\sim$2$\times
10^{-2}\,M_\odot$ with velocities of 3000--4500$\,$km$\,$s$^{-1}$ for
the nickel and 3000--4000$\,$km$\,$s$^{-1}$ for the oxygen. In 
contrast, the smaller finger-like structures move with
1500--3000$\,$km$\,$s$^{-1}$ and contain a total mass of
$\sim$4$\times 10^{-2}\,M_\odot$.

It is amazing that in our 3D simulations the fastest of the metal
clumps have achieved to overtake most of the hydrogen
at the time the shock breaks out (at roughly 8000$\,$s).
The nickel and oxygen mixing into the hydrogen
envelope is clearly more efficient than even in the globally
asymmetric explosions studied by Kifonidis et al.\ (2006), where 
at 10000$\,$s the metals were seen to be distributed only up to an 
enclosed mass of about 10$\,M_\odot$ (see figure 6 there), although
the peak metal velocities were roughly the same at $\sim$300$\,$s.

We refrain here from drawing far reaching conclusions on the 
observational consequences of this finding, e.g.\ concerning the early 
visibility of X-ray and $\gamma$-ray signatures of such strong mixing.
While we consider the 2D/3D differences for the same explosion model
and the same and fixed numerical resolution of our first explorative 
study as enlightening, we think that more simulations with different 
explosion energies, different progenitor stars, and in particular
also with finer grid zoning are desirable before the results
can be interpreted quantitatively with respect to observations of
supernovae.

\section{A simple model for clump propagation}
\label{sec:clumpprop}
In the following we will present a simple analysis that provides
support for our hypothesis that the big discrepancies found in 3D 
compared to 2D models are to a large extent the consequence of 
geometry-dependent differences of
the propagation and deceleration of objects (``clumps'') under the
influence of drag forces in the stellar medium of the helium layer.

\subsection{Time-independent toy model}
A first qualitative understanding can be obtained by considering
objects moving in a stellar medium with a time-independent density
$\rho(r)$. We assume that the clump mass $m = \rho_\mathrm{c}
V_\mathrm{c}$ remains constant and the volume $V_\mathrm{c}$ and
density $\rho_\mathrm{c}$ of the clumps adjust to the local
environmental conditions accordingly.  With a velocity $v$ relative to
the surrounding gas the equation of motion of the clumps at a radius
$r$ can be written as
\begin{equation}
m {{\mathrm{d}}v(r)\over {\mathrm{d}}t}\,=\, {1\over 2}
\, m {{\mathrm{d}}v^2(r)\over {\mathrm{d}}r} \,=\, 
-\,F_{\mathrm{drag}}(r)\,,
\label{eq:motion}
\end{equation}
where the drag force for high Reynolds numbers ($\mathrm{Re} = \rho v
V_\mathrm{c}(A_\mathrm{c} \mu)^{-1} \gg 1$ with $\mu$ being the dynamical
viscosity and $A_\mathrm{c}$ the area the object presents to the flow)
is given by
\begin{equation}
F_{\mathrm{drag}}(r) \,=\, {1\over 2}\, C A_\mathrm{c}(r) 
\rho(r) v^2(r) \,.
\label{eq:fdrag}
\end{equation}
The drag coefficient $C$ is a dimensionless parameter of order unity,
which depends on the shape and the surface structure of the moving
object.  Equation~(\ref{eq:motion}) with Eq.~(\ref{eq:fdrag}) has an
analytic solution for the velocity as a function of distance:
\begin{equation}
{v^2(r)\over v^2_0}\,=\,
\exp  \left ( -\,{C\over m}\int_{r_0}^r {\mathrm{d}}r'\,
A_\mathrm{c}(r') \rho(r') \right )\,\equiv\,{\mathrm e}^{-2 f(r)}\,,
\label{eq:velocity}
\end{equation}
when $v_0$ is the initial velocity of the object at radial position
$r_0$.  The exponential decay of the initial velocity due to the
influence of the drag force depends on the density of the medium
through which the object propagates, on the drag coefficient, the mass
of the object, and the area it presents to the flow.

In the following, we will discuss how the term in the exponent depends
on the dimensionality of the simulation. Because of the assumed axial
symmetry, in 2D all objects have a toroidal form. Let us consider in
2D an ideal torus with radius $r_\mathrm{t}$, whose center has a
distance $r$ from the origin of the stellar grid and a distance $r
\sin\theta$ from the symmetry axis, and compare its outward
propagation in radial direction with that of a sphere with radius
$r_\mathrm{s}$ in 3D also at radial distance $r$ from the center.

These objects, torus and sphere, are assumed to have a fixed shape but
their size can vary as they move in radial direction from the initial
position $r_0$ to $r$. In the beginning, the torus radius is
$r_{{\mathrm t},0}$, and the spherical blob has a radius 
$r_{{\mathrm s},0}$.  In order to compare the propagation of
structures in 3D and 2D as in the hydrodynamical simulations, we
suppose that the torus and sphere have initially the same cross
section and thus radius: $r_{{\mathrm s},0} = r_{{\mathrm t},0}$.
Since we assume the same density inside the objects, 
$\rho_{{\mathrm t},0} = \rho_{{\mathrm s},0}$, 
but the torus volume
$V_{{\mathrm t},0} = 2\pi^2 r_{{\mathrm t},0}^2 r_0 \sin\theta$ 
in general differs from the volume of the sphere, 
$V_{{\mathrm s},0} = {4\pi\over 3} r_{{\mathrm s},0}^3$, 
the masses of torus and sphere are also unequal: $m_{\mathrm t} \neq
m_{\mathrm s}$. Besides these different masses, the two objects also
present different areas to the flow, 
$A_{{\mathrm t},0} = 4\pi r_{{\mathrm t},0} r_0 \sin\theta$ 
and
$A_{{\mathrm s},0} = \pi r_{{\mathrm s},0}^2$.

Moving in radial direction, the clumps are assumed to maintain
pressure equilibrium with the surrounding medium.  As a consequence,
their density and therefore their volume and cross section will
change. For the torus one has the scaling relations
$V_{\mathrm t} = V_{{\mathrm t},0} (r_{\mathrm t}/r_{{\mathrm t},0})^2(r/r_0)$ 
and 
$A_{\mathrm t} = A_{{\mathrm t},0}(r_{\mathrm t}/r_{{\mathrm t},0}) (r/r_0)$, 
while for the sphere they are 
$V_{\mathrm s} = V_{{\mathrm s},0}(r_{\mathrm s}/r_{{\mathrm s},0})^3$
and 
$A_{\mathrm s} = A_{{\mathrm s},0}(r_{\mathrm s}/r_{{\mathrm s},0})^2$.
Using this and the equality of the initial density and radii
of torus and sphere, the arguments in the exponential function of
Eq.~(\ref{eq:velocity}) can be compared for the two geometries to
yield the ratio: 
\begin{equation}
{f_{\mathrm t}(r)\over f_{\mathrm s}(r)}\,=\,
{C_{\mathrm t}\over C_{\mathrm s}}{8\over 3\pi}
\left [ \int_{r_0}^r \! {\mathrm d}r'\,\rho(r') \,{r_{\mathrm t} \over
r_{{\mathrm t},0}}\, {r'\over r_0} \right ]
\left [ \int_{r_0}^r \! {\mathrm d}r'\,\rho(r') \,\left ( {r_{\mathrm s} 
\over r_{{\mathrm s},0}} \right )^{\! 2} \right ]^{-1} . 
\label{eq:exponentratio1}
\end{equation}
In this equation the radii of the torus, $r_{\mathrm t}$, and of the
sphere, $r_{\mathrm s}$, in general depend on the distance $r$ from
the grid center, if the clump propagates through a stellar medium with
a pressure gradient. It should be noted that the ratio $f_{\mathrm
  t}(r)/f_{\mathrm s}(r)$ is nearly independent of both the mass
difference and the initial difference of the torus and sphere areas
perpendicular to the radial direction (the numerical factor $8/(3\pi)$
is very close to unity). Instead, the deceleration of the motion by
drag forces depends crucially on the evolution of the areas as
reflected by the radius-dependent functions in the integrands.

We assume that the pressure in the clump (torus or sphere) scales with
the density in the clump as
$P_{\mathrm c}\propto \rho_{\mathrm c}^m \propto V_{\mathrm c}^{-m}$, 
and that the pressure in the stellar environment follows a radial
profile $P(r)\propto r^{-n}$. Using this and the requirement of
pressure equilibrium, we find that the radii of torus and sphere
change with $r$ according to
\begin{equation}
{r_{\mathrm t}\over r_{{\mathrm t},0}} \,=\,
\left ( {r\over r_0}\right )^{\! {1\over 2}{n\over m} - {1\over 2}} 
\,,\quad\ \ 
{r_{\mathrm s}\over r_{{\mathrm s},0}} \,=\,
\left ( {r\over r_0}\right )^{\! n\over 3m} .
\label{eq:scalings}
\end{equation}

Plugging Eq.~(\ref{eq:scalings}) into Eq.~(\ref{eq:exponentratio1}),
we obtain
\begin{eqnarray}
{f_{\mathrm t}(r)\over f_{\mathrm s}(r)}\,=\,
{C_{\mathrm t}\over C_{\mathrm s}}{8\over 3\pi} &&
\left [ \int_{r_0}^r \! {\mathrm d}r'\,\rho(r') \,
\left ( {r'\over r_0} \right )^{\! {1\over 2}{n\over m} + {1\over 2}}
\right ]   \cr
\times &&
\left [ \int_{r_0}^r \! {\mathrm d}r'\,\rho(r') \,\left ( 
{r'\over r_0} \right )^{\! 2n\over 3m} \right ]^{-1} .
\label{eq:exponentratio2}
\end{eqnarray}
The torus area perpendicular to the radial direction grows faster than
that of the sphere if $n < 3m$, in which case the integral in the
numerator increases more rapidly with $r$ than the one in the
denominator.  This difference is a consequence of the fact that any
radial motion leads to a larger distance of the torus from the polar
axis of the grid. This stretches the torus and causes an adjustment of
its cross section according to pressure equilibrium.  Since in
addition to the area effect also the drag coefficient of the toroidal
object is likely to be larger than that of the sphere, $C_{\mathrm t}
\ga C_{\mathrm s}$, the motion of the torus is clearly more strongly
damped by the drag than that of the sphere if $n < 3m$.

In the supernova the clumps propagate through the medium of the dense
shell of shock-accelerated progenitor matter. In this shell the
density and pressure gradients are nearly flat, i.e., $\rho \sim
{\mathrm{const}}$ and $n\sim 0$ (see, e.g., \citet{nomoto:1994a},
figs.\,31 and 32). This case is particularly suitable to illustrate
the inadequacy of axisymmetric toroidal clumps in 2D instead of
spherical clumps in 3D. While the latter propagate through the
homogeneous medium without any change of their area perpendicular to
the direction of the motion, the tori are unavoidably stretched as
they reach larger radii $r$, which leads to an increase of the
cross-sectional area they present to the flow.
Equation~(\ref{eq:exponentratio2}) yields:
\begin{equation}
{f_{\mathrm t}(r)\over f_{\mathrm s}(r)}
\left ( {C_{\mathrm t}\over C_{\mathrm s}}{8\over 3\pi} 
\right )^{\! -1} \,=\,
{2\over 3}\,
\left [ \left ({r\over r_0}\right )^{\! {3\over 2}} - 1 \right ]\,
\left [ {r\over r_0} -1 \right ]^{-1} ,
\label{eq:exponentratio3}
\end{equation}
which is larger than unity for $r > r_0$. A stronger deceleration 
of the toroidal clump is thus caused by the artificial constraints 
associated with the axisymmetry of all structures in the 2D geometry.

\subsection{Time-dependent toy model}
In a second step we intend to obtain a more quantitative insight by
considering the clump propagation in the dense helium layer that forms
after the passage of the outgoing shock through the He/H interface
\citep[for details, see][]{kifonidis:2003a}. The interaction with this
shell was identified as the main reason why metal-carrying rising
Rayleigh-Taylor structures were unable to penetrate into the H-shell
in the 2D simulations of \citet{kifonidis:2003a}. Here we will
integrate the equation of motion for spherical clumps decelerated by
the drag force in the 3D environment (in contrast to the 3D case of a
spherical clump, the equation of motion of a 2D toroidal object defies
an analytic solution due to the explicit dependence of the clump
volume and cross-sectional area on the radial coordinate position
$r$).

We assume that the He layer is homogeneous but has a time-dependent
density $\rho(t)$. This is an acceptable simplification because
numerical simulations show that the radial density variations in this
layer are much smaller than the time-dependent decrease of the density
due to the expansion of the star during the first three hours of the
explosion. In the 3D case this leaves us just with a time-dependence
of the drag force, $F_{\mathrm{drag}} = F_{\mathrm{drag}}(t)$, and
thus
\begin{equation}
m {{\mathrm{d}}v(r)\over {\mathrm{d}}t}\,=\, -{1\over 2}
\, C_\mathrm{s} A_\mathrm{s}(t)\rho(t) v^2(t)  \,.
\label{eq:motion2}
\end{equation}
Again we consider $m = \mathrm{const}$ and $C_\mathrm{s} = \mathrm{const}$
and $v(t)$ as the clump velocity relative to the surrounding medium.  We
further assume that the density of the helium layer evolves like
$\rho(t) = \rho_0 (t_0/t)^k$ and the pressure and density are linked
by $P(t) = K\rho^\gamma (t)$. Using the same pressure-density relation
for the medium inside the clumps, $P_\mathrm{s} = K_\mathrm{s}
\rho_\mathrm{s}^\gamma$, but with a different constant $K_\mathrm{s}
\neq K$ (corresponding to an entropy of the clumps different from that
of the surroundings), the request of pressure equilibrium between
clumps and environment allows us to calculate the time evolution of
the clump radius,
\begin{equation}
r_\mathrm{s}(t) \,=\, r_{\mathrm{s},0} \left ({K_\mathrm{s}\over
K}\right )^{\! {1\over 3\gamma}} \left ( {\rho_{\mathrm{s},0}\over
\rho_0 }\right )^{\! {1\over 3}} \left ( {t\over t_0}\right 
)^{\! {k\over 3}} ,
\label{eq:clumprt}
\end{equation}
and therefore the time-dependent cross-sectional area that the clumps
present to the surrounding medium:
$A_\mathrm{s}(t) = A_{\mathrm{s},0}(r/r_{\mathrm{s},0})^2 \equiv A^\ast
(t/t_0)^{2k/3}$ with constant $A^\ast$. 
Inserting this and $\rho(t)$ into Eq~(\ref{eq:motion2}), we can
integrate the resulting differential equation by separation of
variables. This yields the time evolution of the velocity (for $k \neq
3$) as
\begin{equation}
v(t)\,=\, v_0 \,\left\lbrace 
1\,+\, {C_\mathrm{s}A^\ast \over 2m} v_0t_0\rho_0 {1\over 1-k/3}
\left [ \left ({t\over t_0}\right )^{\! 1-k/3} - 1 \right ]
\right\rbrace^{-1} .
\label{eq:tsolution1}
\end{equation}
We specialize now to $k = 2$ and $\gamma = {4\over 3}$, which is in
reasonable agreement with the simulations \citep{kifonidis:2003a}.
Rewriting the constant factor of the second term in the denominator by
using the expressions for clump area and volume, we end up with
\begin{equation}
v(t)\,=\, v_0 \,\left\lbrace 
1\,+\, {9\over 8}\left (\! {4\pi\over 3}\! \right )^{\!\! {1\over 3}}\!\!
\left ({K_{\mathrm s}\over K}\right )^{\!\! {1\over 2}} \! C_\mathrm{s}
v_0t_0\left ( {\rho_0\over m}\right )^{\!\! {1\over 3}} \!\! 
\left [ \left ({t\over t_0}\right )^{\!\! {1\over 3}}\!\! - 1 \right ]
\right\rbrace^{-1} \!\!\! .
\label{eq:tsolution2}
\end{equation}
The fastest clumps, which are also the most massive ones with $m\sim
4\times 10^{-3}\,M_\odot$ in our 3D simulations, enter the dense
helium shell with a velocity of $v_0 \sim
2000$--4000$\,$km$\,$s$^{-1}$ (relative to the ambient medium) at
$t_0\sim 300\,$s. At this time the density of the medium in the shell
is $\rho_0\sim 0.1\,$g$\,$cm$^{-3}$.  The clumps typically have
roughly three times lower entropy than the medium in the helium shell
so that $K_\mathrm{s}/K \sim 1/3$ (most of the numerical values used
here can be verified by closely inspecting the plots in
\citet{kifonidis:2003a}). Adopting for the drag coefficient plausible
numbers\footnote{Unfortunately, no information is available in the
  literature for numbers that apply to the problem considered here.}
of $C_\mathrm{s}\sim 0.05\,\ldots\,0.2$, we obtain at $t = 9000\,$s
velocities of $v \sim 1200$--3100$\,$km$\,$s$^{-1}$ for the clump
motion through the helium material, corresponding to a deceleration by
roughly a factor of 2.  This means that the fastest objects are
expected to still speed with $\sim\,$2400--4400$\,$km$\,$s$^{-1}$
relative to an observer in the lab frame.  This is in good agreement
with our hydrodynamical simulations and can thus be interpreted as
support of our picture of the clump propagation under the influence of
the drag force\footnote{Note that the numerical result for the final
  velocity varies relatively little when different combinations of
  values for $v_0$, $\rho_0$, and $t_0$ within reasonable ranges are
  adopted.}. In contrast, in 2D models the maximum clump velocities
are reduced to slightly more than 1500$\,$km$\,$s$^{-1}$ in the lab
frame (see Figs.~\ref{fig:3D2Dvsvel}, \ref{fig:HNivsvel} 
and \citet{kifonidis:2003a}), which is equal to the
expansion velocity of the He-shell matter so that the clumps get stuck
in their surroundings and are unable to penetrate into the hydrogen
envelope (Fig.~\ref{fig:3D2Dvsmass}).

\section{Summary and conclusions}
We have presented the first 3D simulations that followed the
development of mixing instabilities in a supernova over a timescale of
hours after the explosion was initiated by neutrino energy
deposition. The considered progenitor was a 15.5$\,M_\odot$ blue
supergiant star, which had also been adopted for the works of
\citet{kifonidis:2003a, kifonidis:2006a}. Our long-time runs were
started at roughly 0.5$\,$s after core bounce from 3D models provided
by \citet{scheck:2007a}, who triggered the blast by suitably chosen
neutrino luminosities imposed at the retreating inner grid boundary
that replaced the high-density core of the nascent, contracting
neutron star. The supernova runs we focussed our discussion on had an
explosion energy of $10^{51}\,$ergs.  Our particular goal was a close
comparison of the growth of mixing instabilities at the composition
interfaces of the exploding star in 3D and 2D calculations. For the
latter, we investigated a set of selected meridional slices of the
initial 3D explosion models.

Our comparison revealed that the asymptotic velocities of metal-rich
clumps are much higher in 3D than in the corresponding 2D cases. In
the latter the iron-group and oxygen carrying dense fingers get stuck
in the massive, dense helium shell (``wall'') that forms below the
base of the hydrogen envelope, and stay comoving with the helium matter
there at a velocity of $\la$1500$\,$km$\,$s$^{-1}$, a fact that was
pointed out by \citet{kifonidis:2003a}, whose results are in 
agreement with our 2D simulations. In contrast, in the 3D runs we
obtained ``nickel'' ($\equiv\,$elements
of the iron group plus silicon)\footnote{We note again
that in our present models without nuclear burning a small region of 
silicon in the collapsing stellar core was lumped together with the
iron-group nuclei to what we called ``nickel''. Silicon was therefore
not traced separately as a nucleosynthetic constituent of the ejecta.}
and oxygen (plus neon and magnesium) bullets
propagating at maximum velocities of
4500$\,$km$\,$s$^{-1}$ and a large fraction ($\sim$10--20\%) of the
metal core of the progenitor star was seen to reach velocities
of more than 2000$\,$km$\,$s$^{-1}$. The most extended and fastest
Rayleigh-Taylor structures, some of which have a mass of several
$10^{-3}\,M_\odot$, were found to contain mostly nickel. These nickel
``bullets'' expand significantly more rapidly than the longest fingers
with dominant or appreciable oxygen content, which are smaller on
average but much more numerous than the large nickel
features. Iron-group nuclei, neon, and magnesium are thus carried far
into the hydrogen layer. They move well ahead of oxygen-rich knots and
both iron and oxygen overtake the material from the carbon layer in
the ejecta. The onion-shell stratification of the progenitor is thus
partially turned over during the explosion.

Besides strong mixing of heavy-elements into the hydrogen envelope, we
also found about 0.086$\,M_\odot$ of hydrogen being transported 
deep into the metal core of the star to regions with velocities of 
$<$1000$\,$km$\,$s$^{-1}$, $\sim$0.034$\,M_\odot$ to
an enclosed mass of $M(r) < 2\,M_\odot$, and $\sim$0.19$\,M_\odot$ to
$M(r) < 4\,M_\odot$. Although this result seems to be roughly
compatible with observational constraints based on the lightcurve 
shape of SN~1987A (see Blinnikov~1999 and Kifonidis et al.\ 2006), 
we refrain from a more detailed comparison and too far reaching
conclusions, because neither the considered progenitor star
is very suitable to compare with SN~1987A, nor were the explosion
parameters (mass cut, nickel mass, explosion energy) adjusted to 
be optimally consistent with the observations of this supernova.
Moreover, the exact amount of inward hydrogen mixing must be expected
to be sensitive to the numerical resolution and thus at least
convergence tests with better spatial zoning are required.

\citet{kifonidis:2006a} obtained similar results on the basis of
2D explosion models where a supernova shock with a large global
asphericity triggered strong Richtmyer-Meshkov instability at its
passage through the composition interfaces, in particular also
at the helium/hydrogen boundary. Such an effect, however,
does not play any significant role in the present 3D simulations, 
because the shock starts out with much less initial deformation and
is close to spherical when it crosses from the He into the H shell.
Nevertheless, the peak velocities of the metals in our 3D simulations
are even higher than those found by \citet{kifonidis:2006a}, and 
correspondingly the fastest nickel and oxygen clumps are able to
penetrate even farther into the hydrogen envelope.

A closer comparison showed that a small fraction of the metal core
receives velocities of more than 6000$\,$km$\,$s$^{-1}$ initially,
which is significantly higher than in our 2D runs (but similar to
the peak metal velocities seen by \citet{kifonidis:2006a}). These
clumps penetrate into the hydrogen envelope faster than the helium
wall builds up. They achieve to retain a speed of more than 
4000$\,$km$\,$s$^{-1}$ at the time of shock breakout and experience
much less deceleration than in the 2D models of \citet{kifonidis:2006a}.
Moreover, a larger fraction of the core metals start out with 
velocities of about 4000$\,$km$\,$s$^{-1}$, which is of the order
of the speed of the fastest clumps in our 2D models, but again the 
deceleration of these dense bullets is less strong than in the 2D 
case. This suggests that the final differences can be attributed only 
partly to 2D/3D differences in the growth rate of perturbations at
the composition interfaces, which were pointed out by \citet{kane:2000a}.
Instead, the discrepancies between 3D and 2D results that develop when
the majority of the metal-rich clumps begins to enter the dense helium
wall below the He/H boundary after $\sim$1000$\,$s,
and the increase of these discrepancies over the first
1--2 hours after the onset of the explosion while the clunps make their
way through the helium layer, require a different explanation.

We therefore hypothesized that these differences are a consequence
of differences in the drag forces of the surrounding medium, which
affect the propagation of the metal-rich clumps and lead to their
deceleration as they plow through the helium layer. By means of a
simple analytic model we could demonstrate that the different geometry
of the bullets in 2D and 3D --- toroidal (axisymmetric) structures versus
quasi-spherical bubbles --- can explain the differences observed in
our simulations. Making reasonable assumptions in a toy model
description of the conditions in the exploding star and using
plausible values of the drag coefficient, we could come up with
numbers that are in the ballpark of the asymptotic nickel clump
velocities in the hydrogen shell found in our 3D hydrodynamic models.

In this context experimentally and numerically established data for
the drag coefficient in dependence of the clump geometry would have
been highly desirable but could not be found in the literature. A
determination of corresponding values may be a valuable goal for
future investigations, e.g.\ at the National Ignition Facility (NIF)
at Livermore.  It must be suspected that the drag force acting on the
dense metal clumps is not independent of the grid resolution and
insufficient zoning could lead to an underestimation of drag effects
on the clump propagation.  Without applying any adaptive mesh
refinement (AMR) techniques, the 1200 radial zones and 1$^\circ$
resolution in azimuthal and 0.935$^\circ$ in lateral direction were
the best we could allow for as a compromise between accuracy and
CPU-time requirements. Although our 2D results are in reasonably good
agreement with the 2D AMR calculations performed by
\citet{kifonidis:2003a, kifonidis:2006a}, we still think that
simulations with varied mesh size or an AMR code would be useful to
confirm our findings in 3D.

In this pilot study we were aiming at a first analysis of fundamental
differences between mixing instabilities during the long-time
evolution of supernova explosions in two and three dimensions rather
than attempting to match the observational appearance of special
events like, e.g., SN~1987A (although we used this supernova as the
best studied one for our motivation and also referred to it when our
findings agreed with basic features that SN~1987A is likely to have in
common with a wider class of cases). On the basis of our results,
which appear suitable to account for the strong inward mixing of
hydrogen and outward mixing of metals that is needed to explain the
smoothness of the light curve as well as the early appearance of X-ray
and gamma-ray emission of SN~1987A, we therefore do not conclude that
all observational properties of this supernova can be reproduced by
a relatively weakly aspherical initiation of the explosion. More
simulations are needed to clarify this point, in particular by
employing a stellar model that is compatible with all knowledge
acquired about the progenitor structure of SN~1987A during the past
20 years. 

While we think that the 2D/3D differences found in our investigation
are likely to be generic, the extent of the radial mixing, the size
and mass distribution of the metal fingers, their velocities and 
spatial distribution, and the composition of individual clumps
must be expected to depend strongly on the structure of the 
progenitor star, the explosion energy, and the initial asymmetry of
the blast. Having
advanced the modeling of mixing instabilities in supernova explosions
to the third dimension, we therefore plan to direct our focus to the
exploration of a wider variety of progenitor stars with the goal to
establish a link between explosion models and observations.  Besides
SN~1987A, the geometry of the Cassiopeia~A remnant has been studied
observationally in very much detail (e.g., \citet{hwang:2004a,
  fesen:2006a, fesen:2006b}, and references therein). It exhibits
properties such as the radial turnover of the metal core that are
compatible with the findings of this work, and a directional asymmetry
that cannot be accounted for by models constrained to axial symmetry.
Therefore not only the differences of 3D versus 2D results reported in
this paper but also observational properties of supernovae and
supernova remnants point to the indispensability of modeling the
explosions in three dimensions. An interesting question certainly
is, whether the jet-like, silicon-rich features observed in Cas~A could 
be morphologically linked to the structures seen as the fastest and
most massive Rayleigh-Taylor fingers in our simulations. Although our
results appear suggestive in this respect, a convincing answer requires
investigations of progenitors and explosion parameters fitting this
remnant as well as a more complete consideration of the nuclear
composition of the ejecta.

\acknowledgements 
We thank L.~Scheck for providing us with his 3D explosion models as
initial data for our studies, V.~Utrobin for sharing with us his
insight into modelling observations of SN\,1987A, and M.~Rampp 
(Rechenzentrum Garching) for his great assistance in the visualization 
of our 3D data. This work was
supported by the Deutsche Forschungsgemeinschaft through the
Transregional Collaborative Research Centers SFB/TR~27 ``Neutrinos and
Beyond'' and SFB/TR\,7 ``Gravitational Wave Astronomy'', the
Collaborative Research Center SFB-375 ``Astro-Particle Physics'', and
the Cluster of Excellence EXC\,153 ``Origin and Structure of the
Universe'' (\url{http://www.universe-cluster.de}). The computations
were performed on the SGI Altix 4700 of the Leibniz-Rechenzentrum
(LRZ) in Munich.

%%%%%%%%%%%%%%%%%%%%%%%%%%%%%%%%%%%%%%%%%%%%%%%%%%%%%%%%%%%%%
%% Bibliography
%%%%%%%%%%%%%%%%%%%%%%%%%%%%%%%%%%%%%%%%%%%%%%%%%%%%%%%%%%%%%

%%References
\ifthenelse{\boolean{AAorApJ}}
{%%
  \bibliographystyle{aa}
  \bibliography{./biblio.bib}
}
{%%
  \bibliographystyle{apj}
  \bibliography{mnemonic.bib,biblio.bib}
}

\end{document}